\documentclass{jfm}

\usepackage{graphicx}
\usepackage{natbib}
\usepackage{subfigure}
\usepackage{placeins}
\usepackage{amssymb}
\usepackage{amsfonts}
\ifCUPmtlplainloaded \else
  \checkfont{eurm10}
  \iffontfound
    \IfFileExists{upmath.sty}
      {\typeout{^^JFound AMS Euler Roman fonts on the system,
                   using the 'upmath' package.^^J}%
       \usepackage{upmath}}
      {\typeout{^^JFound AMS Euler Roman fonts on the system, but you
                   dont seem to have the}%
       \typeout{'upmath' package installed. JFM.cls can take advantage
                 of these fonts,^^Jif you use 'upmath' package.^^J}%
      }
  \else
  \fi
\fi

\ifCUPmtlplainloaded \else
  \checkfont{msam10}
  \iffontfound
    \IfFileExists{amssymb.sty}
      {\typeout{^^JFound AMS Symbol fonts on the system, using the
                'amssymb' package.^^J}%
       \usepackage{amssymb}%
       \let\le=\leqslant  
         
      }{}
  \fi
\fi

\ifCUPmtlplainloaded \else
  \IfFileExists{amsbsy.sty}
    {\typeout{^^JFound the 'amsbsy' package on the system, using it.^^J}%
     \usepackage{amsbsy}}
    {}
\fi

\newsavebox{\astrutbox}
\sbox{\astrutbox}{\rule[-5pt]{0pt}{20pt}}

\title[Dynamics of line plumes]{Dynamics of line plumes on horizontal
  surfaces in turbulent convection}

\author[G. S. Gunasegarane, and B. A. Puthenveettil]%
{G.\ns S.\ns G\ls U\ls N\ls A\ls S\ls E\ls G\ls A\ls R\ls A\ls N\ls E\ns
\break
\and B\ls A\ls B\ls U\ls R\ls A\ls J\ns A.\ns P\ls U\ls T\ls H\ls E\ls
N\ls V\ls E\ls E\ls T\ls T\ls I\ls L\thanks{apbraj@iitm.ac.in}}

\affiliation{Department of Applied Mechanics, Indian Institute of Technology Madras,
\newline Chennai, 600036, India\\[\affilskip]}

\pubyear{}
\volume{}
\pagerange{}
\date{?; revised ?; accepted ?. - To be entered by editorial office}
\begin{document}

\maketitle

\begin{abstract}
  We study the dynamics of line plumes on the bottom plate in
  turbulent convection over six decades of Rayleigh number
  $(10^5<Ra_w<10^{11})$ and two decades of Prandtl number
  $(0.7<Pr<600)$. From the visualisations of these plumes in a
  horizontal plane close to the plate, we qualitatively identify the
  main dynamics as motion along the plumes, lateral merging of the
  plumes and initiation of the plumes; various other minor types of
  motion also occur along with these main dynamics. The mean velocity
  along the length of the plumes scales as the large scale flow
  velocity, with the fraction of the length of the plumes affected by
  shear increasing with $Ra_w$ as $L_{ps}/L_p\sim Ra_w^{0.04}
  Pr^{-0.1}$. In agreement with \cite{how}, the mean time of
  initiation of the plumes $\overline{t^*}$, scales as the diffusive
  time scale near the plate, $Z_w^2/\alpha$, where $Z_w$ is the
  appropriate length scale near the plate.  Merging occurs in a large
  fraction of the area of the plate, with about 70\% of the length of
  the plumes undergoing merging at $Ra\approx 10^{11}$ and $Sc\approx
  600$.  The fraction of the length of the plumes that undergoes
  merging decreases with increase in $Ra_w$ as, $L_{pm}/L_p \sim
  Ra_w^{-0.04} Pr^{0.1}$; the exponents of $Ra_w$ and $Pr$ being of
  the same magnitude but of opposite sign as that in the relation for
  $L_{ps}/L_p$. Measurements of the velocities of merging of the
  plumes $(V_m)$ show that they merge with a constant velocity during
  their merging cycle.  However, the values of these constant
  velocities depend on the location and the time of measurement, since
  the merging velocities are affected by the local shear. The merging
  velocities at all $Ra_w$ and $Pr$ have a common lognormal
  distribution, but their mean and variance increased with increasing
  $Ra_w$ and decreasing $Pr$.  We quantify the dependence of the mean
  merging velocities $(\overline{V_m})$ on $Ra_w$ and $Pr$, using mass
  and momentum balance of the region between two merging
  plumes. $\overline{V_m}$, which are an order lower than the large
  scale velocities, scale as the average entrainment velocity at the
  sides of the plumes. This implies that $\overline{V_m}$ scales as
  the velocity scale near the plate $\nu/Z_w$.  $Re_H$, the Reynolds
  number interms of $\overline{V_m}$ and the layer height $H$ scales
  as $Ra^{1/3}$, in the same way as the Nusselt number $Nu$; therefore
  $Re_{H}\sim Nu$. These relations imply that $Re_w=
  \overline{V_m}Z_w/\nu$ the Reynolds number near the plate is an
  invariant for a given fluid in turbulent convection.
\end{abstract}

\section{Introduction}\label{sec:Introduction}
In turbulent natural convection over horizontal hot surfaces, rising
sheets of fluid near the plate show complex dynamics of forming and
merging, while being also sheared by the large scale flow. Since these
`line plumes' transport the major portion of the heat from the plate
to the bulk, understanding this dynamics is important to clarify the
phenomenology of flux scaling; improving many technological
applications of convection also depend on such an understanding. In
addition, the dynamics of these plumes strongly influence the velocity
and temperature fields near the plate, whose phenomenology is still
not clear. 
In spite of such importance, the dynamics of these plumes on the
horizontal plate have rarely been investigated quantitatively. In the
present work we quantify various aspects of the dynamics of line
plumes on the bottom horizontal surface in turbulent convection and
then propose scaling laws for the predominant dynamics.

Line plumes are formed from the gravitational
instability\cite*[][]{pera73} of the thin layer of lighter fluid that
forms on the hot horizontal plate.  
Once these line plumes are formed, they rise as sheets while being fed
by the local natural convection boundary layers on both sides of the
sheets, at the same time moving laterally and merging. After a short
distance of rise, they break up into axisymmetric
plumes/thermals 
at heights of the order of 10 times the thermal boundary layer
thickness
\cite*[][]{zhou10:_physic_rayleig}. Line plumes hence collect the
fluid near the plate and transport it into the bulk through the
ejection of thermals. Motions in the bulk group these thermals; the
resultant column of light fluid drives a coherent large scale flow in
the bulk.  The large scale flow in turn creates a shear near the plate
that affects the organisation of these line plumes; the shear also
possibly modifies the local natural convection boundary layers on
either side of the line
plumes~\cite*[][]{puthenveettil05:_plume_rayleig}. In the top views of
these near-plate coherent structures, at any instant, a complex
pattern of lines is seen due to the combined effect of the spatial
nature of the boundary layer instability, the spatially varying
external shear due to the large scale flow and, more importantly, due
to the interaction between these line plumes.

Such complex plume structures, look geometrically quite different at
different Rayleigh numbers ($Ra$) and Prandtl numbers ($Pr$), and have
their mean and integral properties strongly dependent on $Ra$ and
$Pr$. Here, $Ra =g\beta \Delta T H^3/\nu \alpha$ and $Pr=\nu/\alpha$,
with $g$ being the acceleration due to gravity, $\beta$ the
coefficient of thermal expansion, $\Delta T$ the temperature
difference across the fluid layer, $H$ the height of the fluid layer,
$\nu$ the kinematic viscosity and $\alpha=k/\rho C_p$ the thermal
diffusivity with $k$ the thermal conductivity, $\rho$ the density and
$C_p$ the specific heat at constant pressure. To describe the
phenomena near the plate it is also convenient to define a Rayleigh
number based on the temperature drop near the plate $\Delta T_w$, as
\begin{equation}
  \label{eq:nwra}
  Ra_w=g\beta \Delta T_w H^3/\nu \alpha.
\end{equation}
At high $Ra$, since $\Delta T_w=\Delta T/2$, $Ra_w=Ra/2$.  One of the
mean properties of these plume structures at any instant, the mean
plume spacing $\overline{\lambda}$, was shown to be proportional to,
\begin{equation}
  \label{eq:zw}
  Z_w=\left(\frac{\nu\alpha}{g\beta\Delta T_w}\right)^{1/3}=\frac{H}{Ra_w^{1/3}},
\end{equation} 
the appropriate length scale obtained by a balance of buoyant and
diffusive processes near the plate
\cite*[][]{theerthan98:_rayleig,puthenveettil05:_plume_rayleig}. 
Dependence of other mean geometric properties of these line plumes on
$Ra$ have also been obtained empirically by
\cite{zhou10:_physic_rayleig}. \cite{puthenveettil11:_lengt}, herein
after referred as PA11, obtained relations for an integral property of
these plume structures viz. the total length $L_p$, under similar
assumption that natural convection boundary layers, or equivalently
line plumes that are an outcome of these boundary layers, carry most
of the heat from the plate to the bulk. The plume lengths per unit
area $L_p/A\sim1/Z_w$ for any given fluid; a similar $Ra^{1/3}$
dependence were also observed by
\cite{bosbach12:_plume_fragm_bulk_inter_turbul}, eventhough they
interpreted it differently. This result also implied that $L_p H/A
\sim Nu$, the Nusselt number; $Nu= Q/(k\Delta T/H)$, where $Q$ is the
heat flux.

Even though the mean and the integral properties of the plume
structures are strong functions of $Ra$ and $Pr$, the statistical
distributions of the properties of these structures are not.
\cite{puthenveettil05:_plume_rayleig} found that the probability
distribution function (pdf) of the spacings between these line plumes
showed a common log-normal form at different $Ra$ and $Pr$;
\cite{haramina04:_coher_rayleig} have tried to explain such a
distribution. A similar common log-normal distribution has been found
for other geometric properties of these line plumes
by~\cite{zhou10:_physic_rayleig}. 
These plume structures also had a common multifractal spectrum
associated with them over a decade of
$Ra$~\cite*[][]{puthenveettil05:_multif_raylieg}, implying that the
probability of occurrence of the line plumes in any area of the
planform is independent of $Ra$. The common form of the pdfs observed
by various researchers and the common multifractal spectrum point
towards some commonality in the underlying dynamics by which these
structures are formed.

\cite{sprow} was the first to qualitatively describe the dynamics of
these structures at the free surface in evaporative convection; they
identified the transient nature of forming, random motion and plunging
down of line plumes. \cite{husar68:_patter} also made similar
qualitative observations while studying the patterns of line plumes on
a hot horizontal plate with no side walls. The first quantitative
measurements of the horizontal velocities of these line plumes on the
bottom plate, which they considered as waves on the viscous boundary
layer, was by \cite{zocchi90:_coher}. In their visualisations, the line
plumes mostly moved lateral to their length, in the direction of the
large scale flow; the velocity distribution was asymmetric with a long
tail, whose peak was approximately same as the velocity of the large
scale flow. Recently, \cite{hogg13:_reynol_prand} have estimated the
mean and fluctuating horizontal velocities in the convection cell
using spatial correlation of shadowgraph images obtained from the top
view of the cell. Since plumes near the plate are the predominant
structures seen in such images, the obtained velocities are mostly the
plume velocities. \cite{hogg13:_reynol_prand} obtained the scaling of
the mean and the fluctuating velocity to be the same as that of the
large scale flow. The magnitude of the mean velocities were two orders
lower than that of the bulk flow while the magnitude of fluctuating
velocities were of the same order as that of the bulk flow.

Most of the above investigations of the dynamics of these line plumes
on the plate have been qualitative. The few quantitative
investigations fail to describe the whole gamut of motions and
identify the predominant dynamics as that of the large scale flow.
However, as described above, the line plumes have various other
motions other than that due to the large scale flow alone. To the best
of our knowledge, details of the dynamics of line plumes on the hot
plate in turbulent convection are still unexplored. There is no
information available about the various types of motion, the
predominant ones from these types, the relative magnitude of the
predominant motions with the other velocity scales in turbulent
convection and the scaling of the predominant motions with $Ra$ and
$Pr$.  The motion of these line plumes, that have a higher temperature
than in the bulk, contribute to the velocity and temperature profiles
near the plate in turbulent convection. Since various theoretical
expressions for these profiles have been proposed by
\cite{theerthan98:_rayleig}, \cite{shishkina09:_mean_rayleig_benar}
and \cite{ahlers12:_logar_temper_profil_turbul_rayleig_benar_convec},
understanding the quantitative details of the dynamics of line plumes
is crucial to understand the origin of the profiles of temperature and
velocity near the plate. Such profiles near the plate, based on
realistic phenomenology of the dynamics near the plate, could result
in better wall functions for modelling turbulent convection
computationally. More importantly, since line plumes transport most of
the heat from the plate~\cite*[][]{shishkina08:_analy_rayleig},
understanding the dynamics of these structures is necessary to clarify
the phenomenology of heat transport in turbulent convection. Such
knowledge will also be of great use in applications that use natural
convection heat transport like electronic cooling, materials
processing and atmospheric modelling.

In this paper, we study the dynamics of line plumes on the bottom
plate in turbulent convection over horizontal surfaces.  Three
experimental setups that use air ($Pr\approx 0.7$), water ($Pr\approx$
3.6 to 5.3) and concentration driven convection in water ($Sc\approx
600$), are used to achieve a wide range of $Pr$ and Rayleigh numbers
($10^5<Ra_w<10^{11}$). Movies of the evolution of the line plume
structure with time in a horizontal plane grazing the bottom plate are
captured using laser scattering by particles, electrochemical
visualisation and planar laser induced fluorescence (PLIF) for the
three $Pr$ respectively. After identifying the predominant types of
motion from the visualisations as shear, initiation and merging of
plumes, we find scaling laws for the mean shear velocities and the
initiation times. We then focus on describing the mean and statistics
of the merging velocities before proposing scaling laws for the
dependence of merging velocities on $Ra$ and $Pr$.

The paper is organised as follows. After a description of the
experimental setups used for the three $Pr$ in
\S~\ref{sec:Experiments}, we give a qualitative description of the
various types of motion observed on the plate in
\S~\ref{sec:qual-descr-dynam}. In \S~\ref{sec:meas-plume-dynam} we
describe the methodology of our
measurements 
while the results of these measurements are discussed in
\S~\ref{sec:dynamics-line-plumes}. We describe the scaling of
longitudinal velocities in \S~\ref{sec:shear-along-plumes} and the
scaling of initiation times in \S~\ref{sec:initiation-plumes-1}. The
rest of the paper focuses on the dynamics of merging of plumes with
\S~\ref{sec:fract-merg-length} showing that merging occurs in a large
fraction of the total plume length. \S~\ref{sec:vari-merg-veloc} finds
that the plumes merge with constant velocities; the statistics of
merging velocities are the discussed in
\S~\ref{sec:stat-merg-veloc}. Based on the variation of the mean
merging velocities described in \S~\ref{sec:mean-merg-veloc}, we
develop a theory for the scaling of mean merging velocities in
\S~\ref{sec:scaling-mean-merging}. The scaling laws for the mean
merging velocities and the verification of these laws with our
experiments are described in \S~\ref{sec:scaling-lambdaprime}, before
we conclude in \S~\ref{sec:conclusion}.
\section{Experiments}\label{sec:Experiments}
The dynamics of plume structures near the plate were obtained from top
view images in experiments conducted at $Pr=0.74, 3.6$ to $5.3$ and
$602$ for $Ra_{w}$ ranging from $ 10^{5}$ to $10^{11}$. We briefly
discuss the details of the setups and the procedure used for the
conduct of these experiments in this section; details could be found
in~\cite{gunasegarane12:_struc}.
\subsection{$Pr=0.74$}
\label{sec:pr=0.74}
\begin{figure}
  \centering
  \subfigure{\includegraphics[width=0.8\textwidth]{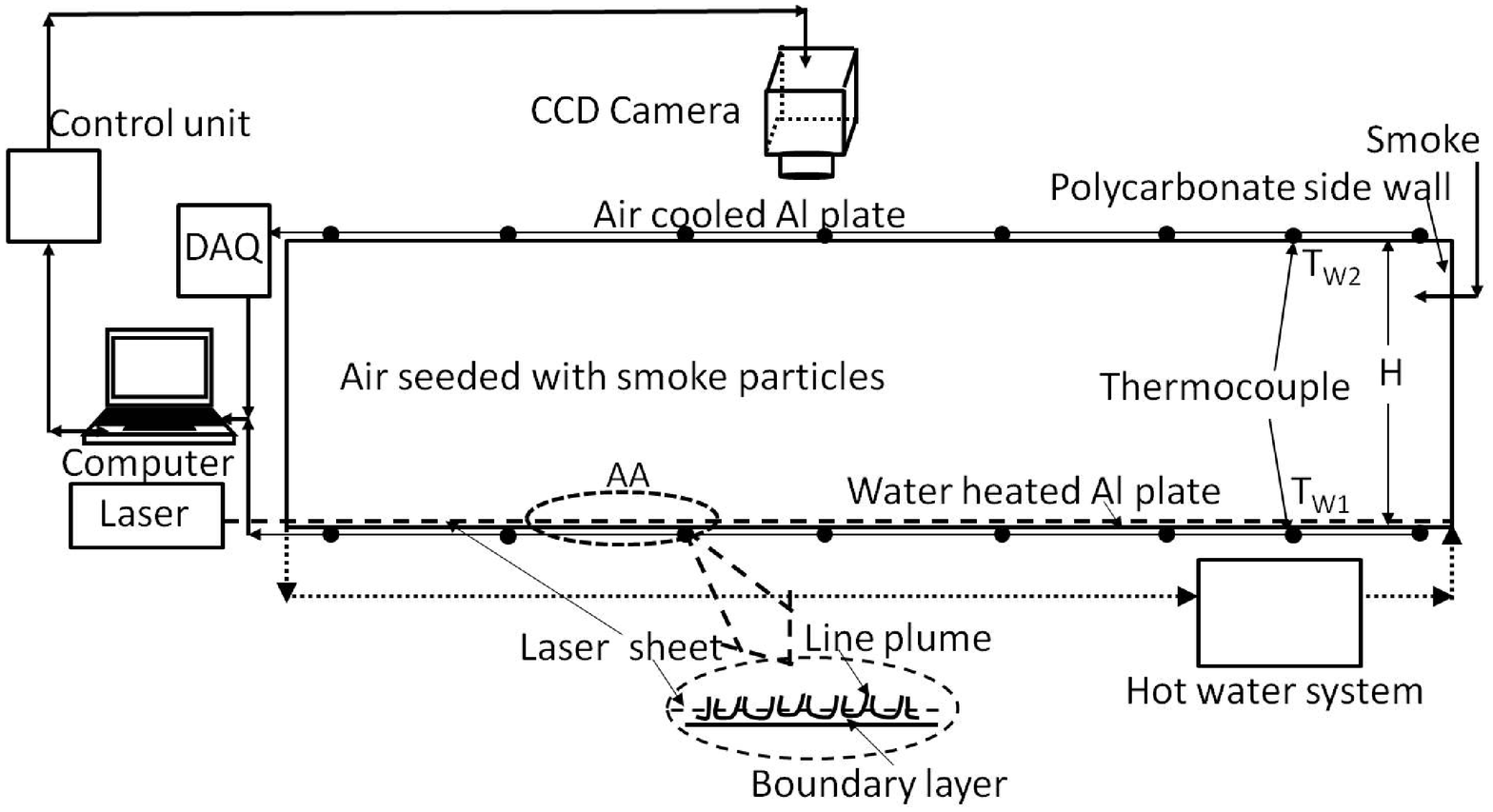}\label{fig:airsetup}}
  \subfigure{\includegraphics[width=0.575\textwidth]{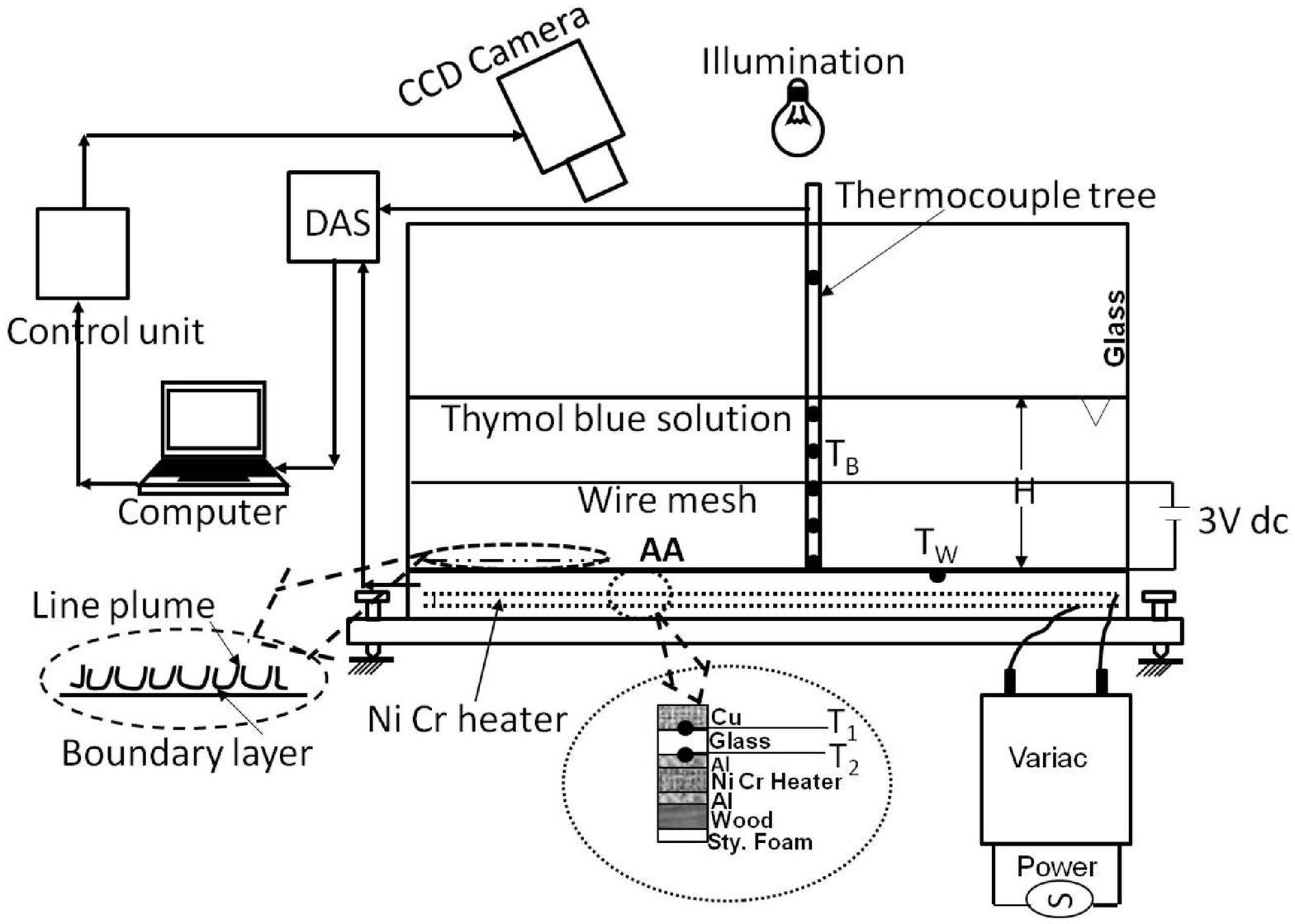} \label{fig:thymsetup}}\hfill
  \subfigure{\includegraphics[width=0.4\textwidth]{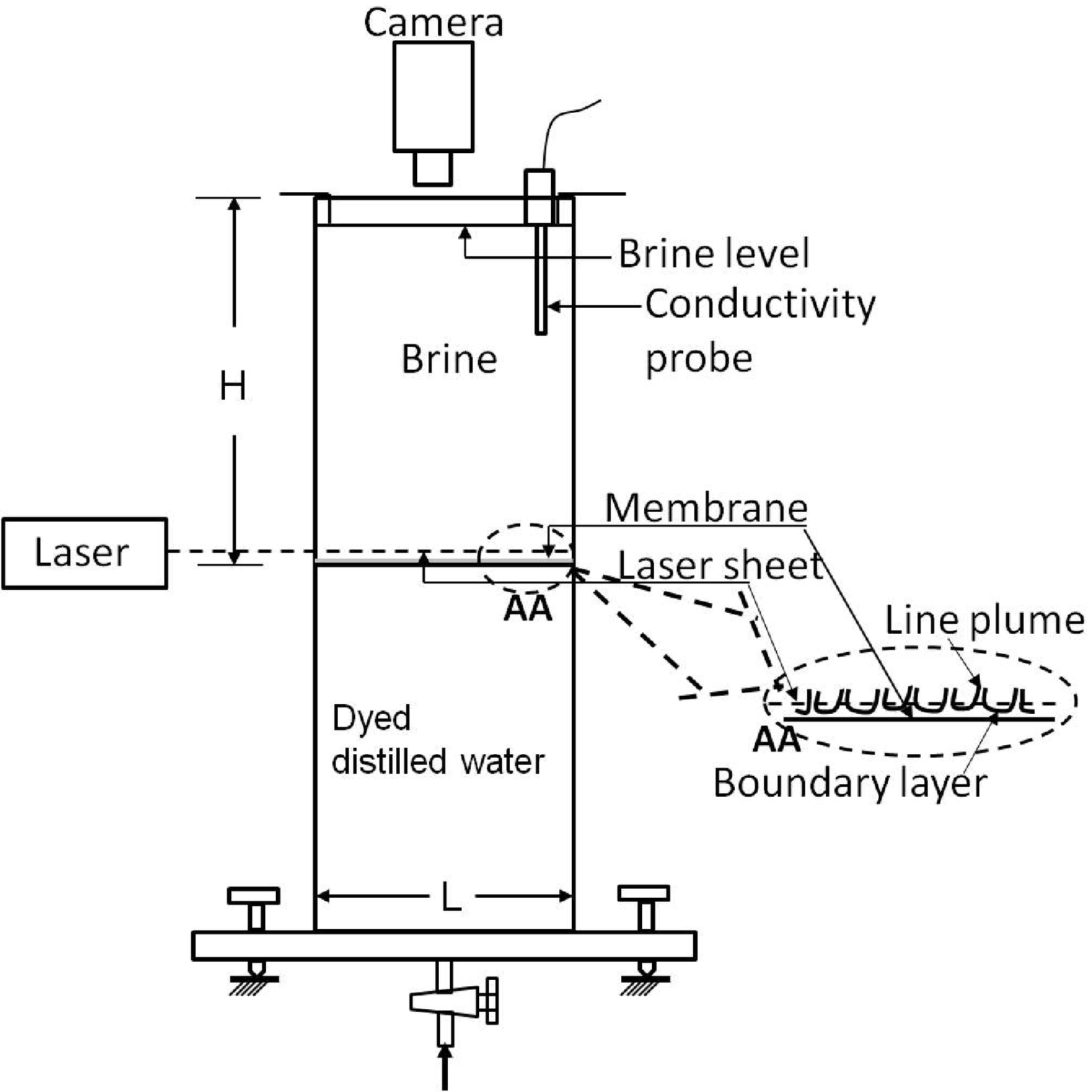} \label{fig:plifsetup}}
  \caption{Schematic of the experimental setups used for studying; (a), steady
    temperature driven convection in air at $Pr=0.74$; (b), unsteady
    temperature driven convection in water at $Pr=3.6$ to $5.3$ and (c), 
    quasi-steady concentration driven convection at $Sc=602$
    \cite*[][]{puthenveettil05:_plume_rayleig}. }
\label{fig:setup2}
\end{figure}

A convection cell of area of cross section $2000$mm$\times500$mm, that
had air confined between top and bottom aluminium plates was used for
visualising the plume structures near the bottom plate at $Pr=0.74$.
The schematic of the convection cell used is shown in
figure~\ref{fig:airsetup}. The aluminium plates were separated by four
transparent polycarbonate side walls of height $500$mm. The bottom
aluminium plate was maintained at a constant temperature using a
temperature controlled water circulating system. The top plate was air
cooled by fans so that a constant mean temperature difference $\Delta
T$ between the plates could be maintained.  $\Delta T$ was determined
from spatial and temporal averaging of the plate temperatures recorded
at 25 locations in each plate using PT100 thermocouples.  The range of
Rayleigh numbers in the experiments was
$1.28\times10^{8}<Ra_{w}<2.54\times10^{8}$, which was obtained by
changing $\Delta T$.  Each experiment was run for approximately $300$
minutes to ensure a steady state condition inside the convection cell
before measurements were taken.  The planforms of plume structures
near the plate were made visible when a horizontal light sheet from a
$532$ nm Nd-Yag laser, grazing the bottom hot plate, was scattered by
the smoke particles mixed with the air inside the convection
cell. Since the plumes have relatively lesser number of smoke
particles, they scatter less light and hence appear as dark lines in a
bright background.

Since the convection cell had a closed opaque top, the top views of
the plume structures near the bottom plate were captured through the
side walls by a CCD camera at 10fps. The perspective errors caused by
this inclined camera axis were removed using a horizontal calibration
plate in the plane of
observation. Figures~\ref{fig:air_noshear_planform} shows one of such
top view of the plume structures near the plate at $Ra_{w}=2.54 \times
10^{8}$, the dark lines in the figure are the top view of the rising
line plumes near the bottom plate. The details of the values of the
parameters used in these experiments are given in
table~\ref{tab:parameter}.
\begin{table}
  \begin{center}
\def~{\hphantom{0}}
  \begin{tabular}{lcccccc}
    $Pr$ or $Sc$ & $H$(mm) & $A_{p}$(mm$^{2}$) & $\Delta T_{w}$or$\Delta C_{w}$   &   $Flux$ 	& $Ra_{w}$ & $\bar{\lambda'}$(mm/s)  \\[3pt]
    0.74 &500&$2500\times500$&10.48 & 19.12 	& $1.287\times 10^{8}$ 	& 16.16\\
    & & & 15.46 				& 31.44 	& $1.766\times 10^{8}$ 	& 19.132\\
    & & & 20.17 				& 44.16 	& $2.183\times 10^{8}$ 	& 20.64\\
    & & & 24.89 				& 57.64		& $2.542\times 10^{8}$ 	& 21.886\\
    5.3	 &50&$300\times300$ & 0.05	& 65 		& $1.31\times 10^{5}$ 	& 0.2608\\
    5.3     & & & 0.15					& 190 		& $3.92\times 10^{5}$ 	& 0.3922\\
    5.2	 & & & 0.26 				& 260		& $7.32\times 10^{5}$ 	& 0.4758\\
    5.1   & & & 0.56 				& 585		& $1.66\times 10^{6}$ 	& 0.5776\\
    5.1	 & & & 1.12					& 1040		& $3.225\times 10^{6}$ 	& 0.7440\\
    4.9	 & & & 2.03 				& 1625		& $6.464\times 10^{6}$ 	& 0.8630\\
    4.7	 & & & 2.53 				& 2340		& $8.49\times 10^{6}$ 	& 0.9496\\
    3.9   & & & 4.65					& 4158		& $2.137\times 10^{7}$ 	& 1.2741\\
    5.3	 &100&$300\times300$ & 0.15	& 50		& $3.14\times 10^{6}$ 	& 0.3866\\
    5.2   & & & 0.85					& 260 		& $1.91\times 10^{7}$ 	& 0.6726\\
    5.1	 & & & 1.19					& 585		& $2.83\times 10^{7}$ 	& 0.7292\\
    4.7   & & & 2.66					& 1625		& $7.145\times 10^{7}$ 	& 1.0448\\

	4.0	 & & & 3.19					& 2340		& $1.11\times 10^{8}$ 	& 1.1044\\
	5.3	 &150&$300\times300$ & 0.21	& 50		& $1.48\times 10^{7}$ 	& 0.4666\\
	5.0     & & & 1.87					& 585		& $1.57\times 10^{8}$ 	& 0.9297\\
	4.7	 & & & 2.92					& 1625		& $2.65\times 10^{8}$ 	& 1.1048\\
	4.0	 & & & 3.26					& 2340		& $3.82\times 10^{8}$ 	& 1.1180\\
    3.6     &200& $300\times300$& 5.41	& 4158		& $1.77\times 10^{9}$ 	& 1.2772\\
	5.3	 &210&$300\times300$ & 0.23	& 50		& $4.46\times 10^{7}$ 	& 0.4498\\
	5.1     & & & 2.45					& 1040 		& $5.39\times 10^{8}$ 	& 0.9818\\
	4.0	 & & & 4.91					& 2340		& $1.58\times 10^{9}$ 	& 1.222\\
      3.6   & & & 5.98					& 4158		& $2.27\times 10^{9}$ 	& 1.3132\\
	602	 &230&$100\times100$ & 1	& 0.021		& $6.39\times 10^{10}$ 	& 0.15\\
	     & & & 2.65					& 0.09		& $1.667\times 10^{11}$ & 0.2094\\
		 & & & 3.21					& 0.1116	& $2.034\times 10^{11}$	& 0.216\\
\end{tabular}
\caption{Values of parameters for the planforms from which the plume
  dynamics were measured. For concentration driven convection at
  $Sc=602$, the driving potential$\Delta C_{w}$ is in gl$^{-1}$ and
  flux in mg cm$^{-2}$ min$^{-1}$ while for other cases it is in
  $^{o}$C and W m$^{-2}$. $H$ is the liquid layer height and $A_{p}$
  is the plate area.}
  \label{tab:parameter}
  \end{center}
\end{table}
\subsection{$Pr=3.6-5.3$}\label{sec:pr=5.2}
The plume structures near the plate in water for $1.31 \times 10^{5}<
Ra_{w} < 2.27 \times 10^{9}$ at $3.6 < Pr< 5.3$, were obtained from
unsteady temperature driven convection experiments conducted in a
convection cell of size $300$mm $\times300$mm $\times250$mm. The
schematic of the set up used in the experiments is shown in
figure~\ref{fig:thymsetup}. The top of the convection cell was kept
open to ambient while the bottom copper plate was used to provide a
constant heat flux into the layer of water above the plate.  The glass
side walls as well as the bottom of the cell were sufficiently
insulated to minimise the heat loss.  The constant heat flux was
provided by a Nichrome wire heater connected to variac, sandwiched
between two aluminium plates in a vertical plate array as shown in
figure~\ref{fig:thymsetup}. The temperature difference
($T_{1}-T_{2}$), measured across the glass plate in this vertical
array of plates was used to estimate the flux supplied to the bottom
plate. This flux was also cross checked with the heat flux supplied by
the nichrome wire, estimated from the resistance of the wire and the
voltage input.  Even though the temperature of the bottom plate $T_w$
and the bulk fluid $T_B$ increased with time, a constant $\Delta
T_{w}=T_w-T_B$ was obtained after about 200 min due to the balance of
influx and efflux of heat.  Since the dynamics near the plate is
solely determined by the driving potential $\Delta T_{w}$ and $Pr$,
the results from the present unsteady experiments could be compared to
that from steady Rayleigh Be'nard Convection (RBC).

When a 5V DC supply is applied across the water layer, the local pH of
water near the cathodic bottom plate changes due to accumulation of
H$^+$ ions. Due to this pH change, the thymol blue dye added to the
water changes its colour near the bottom plate, thereby making the
plume structures near the plate visible~\cite*[][]{baker66}. Since the
change in pH occurs in regions close to the bottom plate alone, only
the plume structures very near the plate are seen by this
technique. The dark lines seen in
figure~\ref{fig:water_merge_no_shear} are the line plumes observed
near the plate using this technique. The parameters of these
experiments are shown in table~\ref{tab:parameter}.  The reader is
referred to \cite{gunasegarane12:_struc} for more details.
\subsection{$Sc=602$}\label{sec:sc=602}
Figure~\ref{fig:plifsetup} shows the schematic of the experimental set
up used by \cite{puthenveettil05:_plume_rayleig} for unsteady
concentration driven convection at $Sc=602$ that occurred in a layer
of brine placed over a layer of water separated by a horizontal
membrane.  The experiments were conducted in a tank of cross section
$100$mm$\times 100$mm for a liquid layer height of $H=230$mm.  The
plume structure shown in figure~\ref{fig:plif_no_shear} was observed
when an Ar-ion laser ($488$nm) sheet, just above the membrane,
intercepted the line plumes that had sodium fluorescein dye in them.
The membrane acted in a way similar to the bottom plate in the
temperature driven convection, allowing only diffusive transport of
mass across it.  Since the time scale of dynamics of line plumes near
the membrane in this system was much lower than the time scale of
variation of flux and the large scale flow, the convection was
quasi-steady (\cite{puthenveettil08:_convec} and
\cite{ramareddy11:_pe}). The results could hence be compared with
steady RBC when the concentration difference $\Delta C_{w}$ in
$Ra_{w}$ is calculated using the concentration difference above/below
the membrane. The reader is referred to
\cite{puthenveettil05:_plume_rayleig}, \cite{puthenveettil08:_convec}
and \cite{ramareddy11:_pe} for more details of the set-up and
procedure.
\section{Qualitative description of dynamics}\label{sec:qual-descr-dynam}
The videos of the topviews of the plume dynamics were studied to
classify the complex dynamics of these plumes into the following four
classes,
\begin{enumerate}
\item longitudinal motion of line plumes,
\item lateral motion of line plumes,
\item initiation of point plumes and
\item miscellaneous motions. 
\end{enumerate}
We now qualitatively describe these classes of motion in this section
before quantifying the first three classes in
\S~\ref{sec:dynamics-line-plumes}.
\subsection{Longitudinal motion of line plumes}
\label{sec:long-moti-line}
 \begin{figure}
   \centering
   \subfigure[]{\includegraphics[width=0.65\textwidth]{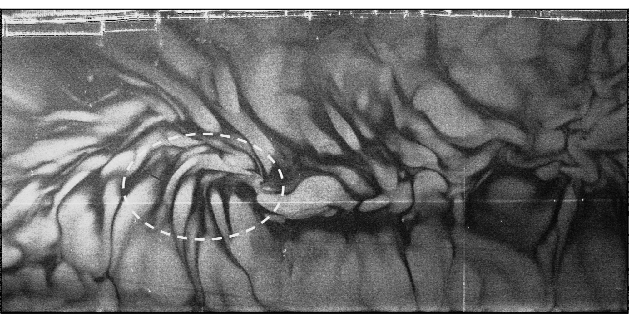}\label{fig:merge_shear_full_planform}}\\
   \subfigure[]{\includegraphics[width=0.225\textwidth]{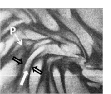}\label{fig:merairshear1}}
   \subfigure[]{\includegraphics[width=0.225\textwidth]{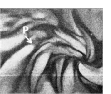}\label{fig:merairshear2}}
   \subfigure[]{\includegraphics[width=0.225\textwidth]{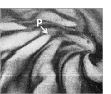}\label{fig:merairshear3}}
   \subfigure[]{\includegraphics[width=0.225\textwidth]{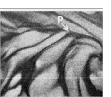}\label{fig:merairshear4}}
   \caption{Dynamics of plumes in a shear dominant region at
     $Pr=0.74$, and $Ra_{w}=2.54 \times 10^{8}$; (a), the
     instantaneous planform in half the cell area of size
     $985$mm$\times 477$mm; (b)-(e), merging sequence in the marked
     region in (a) that has dominant local shear. The filled arrow in
     figure(b) shows the direction of shear, inferred from the motion
     of the point marked as P. The unfilled arrows show the direction
     of merging. The zoomed images are of size $162$mm$\times135$mm at
     times t=0s, 0.5s, 1.0s and 1.5s.}
  \label{fig:merseqairshear}
  \centering
  \subfigure[]{\includegraphics[width=0.4\textwidth]{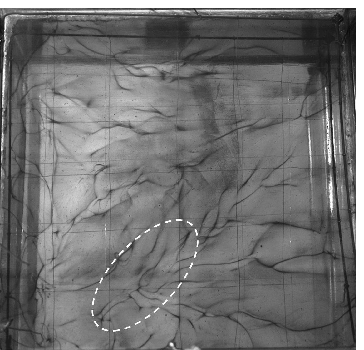}\label{fig:merge_shear_water_full}}\\
  \subfigure[]{\includegraphics[width=0.225\textwidth]{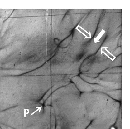}\label{fig:merthymshear1}}
  \subfigure[]{\includegraphics[width=0.225\textwidth]{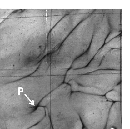}\label{fig:merthymshear2}}
  \subfigure[]{\includegraphics[width=0.225\textwidth]{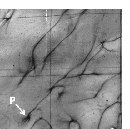}\label{fig:merthymshear3}}
\subfigure[]{\includegraphics[width=0.225\textwidth]{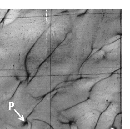}\label{fig:merthymshear4}}
\caption{Dynamics of plumes in a shear dominant region at $Pr=4.7$,
  and $Ra_{w}=2.65 \times 10^{8}$; (a), the instantaneous planform of
  the entire cell area of size $298$mm$\times 287$mm; (b)-(e), merging
  sequence in the marked region in (a) that has dominant local
  shear. The filled arrow in figure(b) shows the direction of shear,
  inferred from the motion of point marked as P. The unfilled arrows
  show the direction of merging. The zoomed images are of size
  $100$mm$\times77$mm at times t=0 s, 5 s, 10 s and 13 s.}
  \label{fig:merseqthymshear}
\end{figure}  
\begin{figure}
  \centering
  \subfigure[]{\includegraphics[width=0.4\textwidth]{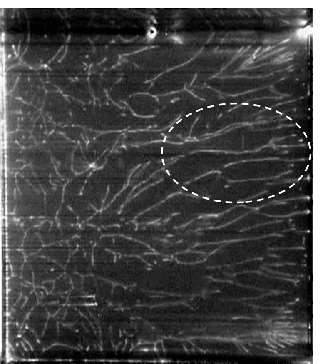}\label{fig:merge_shear_plif_full}}\\
  \subfigure[]{\includegraphics[width=0.225\textwidth]{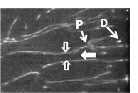}\label{fig:merplifshear1}}
  \subfigure[]{\includegraphics[width=0.225\textwidth]{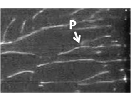}\label{fig:merplifshear2}}
  \subfigure[]{\includegraphics[width=0.225\textwidth]{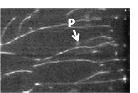}\label{fig:merplifshear3}}
  \subfigure[]{\includegraphics[width=0.225\textwidth]{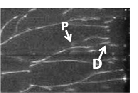}\label{fig:merplifshear4}}
  \caption{Dynamics of plumes observed by
    \cite{puthenveettil05:_plume_rayleig} in a shear dominant region
    at $Pr=602$, and $Ra_{w}=2.034 \times 10^{11}$; (a), the
    instantaneous planform of the entire convection surface of the
    cell of size $87$mm$\times 98$mm; (b)-(e), merging sequence in the
    marked region in (a) that has dominant shear. The filled arrow in
    figure(b) shows the direction of shear, inferred from the motion
    of P. The unfilled arrows show the direction of merging. The
    zoomed images are of size $34$mm$\times23$mm at different merging
    times t=0 s, 2 s, 4 s and 6 s.}
  \label{fig:merseqplifshear}
\end{figure}
There was substantial motion along the lengths of plumes in various
regions of the planforms. Most often, these motions along the length
of the plumes also coincided with the direction of local alignment of
the plumes. Such an instance of motion is seen in
Figures~\ref{fig:merairshear1} to \ref{fig:merairshear4} at
$Ra_w=2.54\times 10^8$ and $Pr=0.74$. The point P in the figure, which
indicates the point of joining of two plumes, moves along the length
of the plumes on both sides of P. The plumes to the bottom left of P
are also aligned along the direction of motion of P. From such a
motion we infer that there is a shear in the planform along the
diagonal connecting bottom left to top right of these images; point P
seems to be at the shear front which moves along this diagonal. This
direction of shear is shown by a filled arrow in
figure~\ref{fig:merairshear1}. The plumes, that are aligned along the
direction of shear, also merge with time, as is obvious by observing
the space between the plumes indicated by the hollow arrows in
figure~\ref{fig:merairshear1}. Such dynamics that include motion of
plume features along their lengths, aligning of plumes along this
direction of motion and the merging of such parallel plumes could also
be seen in figures~\ref{fig:merthymshear1} to \ref{fig:merthymshear4}
at a similar $Ra_w$ of $2.65\times 10^8$ but at a higher $Pr$ of
$4.7$. Comparing the motion of point P in
figures~\ref{fig:merthymshear1} to \ref{fig:merthymshear4} with that
in figures \ref{fig:merairshear1} to \ref{fig:merairshear4} we could
also infer that the velocity of longitudinal motion of the plumes are
substantially less for the higher $Pr$ case. A qualitatively similar
dynamics in the shear dominant region at a much higher $Ra_w$ of $
2.034\times 10^{11}$ and $Sc$ of $600$ is shown in figures
\ref{fig:merplifshear1} to \ref{fig:merplifshear4}. A similar
comparison of the motion of P between figures~\ref{fig:merthymshear1}
to \ref{fig:merthymshear4} with that in \ref{fig:merplifshear1} to
\ref{fig:merplifshear4} does not show a substantial difference in
velocities of longitudinal motion, even though the $Pr$ is much higher
in the latter set of planforms. As we quantify later in
\S~\ref{sec:fract-merg-length}, the strength of longitudinal motion is
also a strong function of $Ra_w$, which being substantially higher for
figure~\ref{fig:merseqplifshear}, offsets the decrease due to a higher
$Pr$.  All the planforms have regions with different strengths of
shear, low shear regions usually occur near to the side walls, with
the high shear region usually occuring at the center.
\subsection{Lateral motion of line plumes}\label{sec:lateral-motion-line}
\begin{figure}
  \centering
  \subfigure[]{\includegraphics[width=0.65\textwidth]{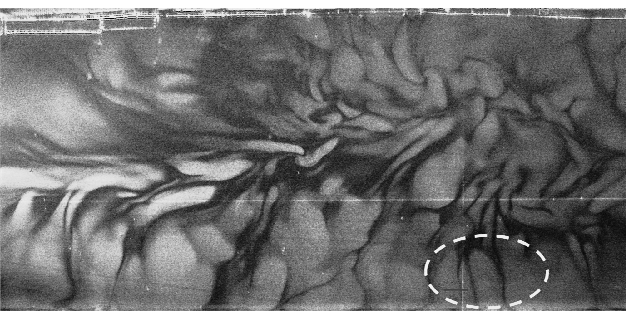}\label{fig:air_noshear_planform}}\hfill\\
  \subfigure[]{\includegraphics[width=0.225\textwidth]{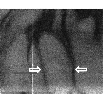}\label{fig:merair1}}\hfill
  \subfigure[]{\includegraphics[width=0.225\textwidth]{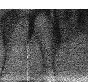}\label{fig:merair2}}\hfill
  \subfigure[]{\includegraphics[width=0.225\textwidth]{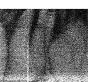}\label{fig:merair3}}\hfill
  \subfigure[]{\includegraphics[width=0.225\textwidth]{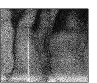}\label{fig:merair4}}
  \caption{Dynamics of plumes in a low shear region at $Pr=0.74$ and
    $Ra_{w}=2.54 \times 10^{8}$; (a), The instantaneous planform in
    half the cell area of size $985$mm$\times 477$mm; (b) to (e),
    merging sequence in a region of size $162$mm$\times135$mm that has
    negligible shear, marked with a dashed ellipse in (a).  The images
    (b) to (e) are separated by $0.4s$.}
\label{fig:merseqairnoshear}
\end{figure}
\begin{figure}
\centering
\subfigure[]{\includegraphics[width=0.4\textwidth]{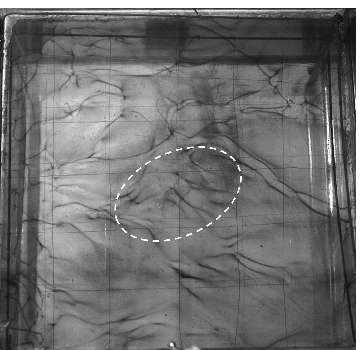}\label{fig:water_merge_no_shear}}\hfill\\
\subfigure[]{\includegraphics[width=0.225\textwidth]{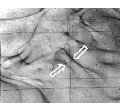}\label{fig:merthym1}}\hfill
\subfigure[]{\includegraphics[width=0.225\textwidth]{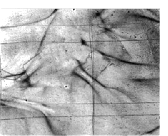}\label{fig:merthym2}}\hfill
\subfigure[]{\includegraphics[width=0.225\textwidth]{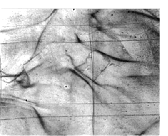}\label{fig:merthym3}}\hfill
\subfigure[]{\includegraphics[width=0.225\textwidth]{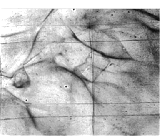}\label{fig:merthym4}}
\caption{Dynamics of plumes in a low shear region at $Pr=4.7$, and
  $Ra_{w}=2.65 \times 10^{8}$; (a), the instantaneous planform of the
  entire cell area of size $298$mm $\times 287$mm; (b) - (e), merging
  sequence in the marked region in (a) with negligible shear; The
  images are of size $100$mm $\times77$mm at times $t=0s,1.5s, 3s$ and
  $5s$ respectively.}
  \label{fig:merseqthymnoshear}
\end{figure} 
\begin{figure}
\centering
\subfigure[]{\includegraphics[width=0.4\textwidth]{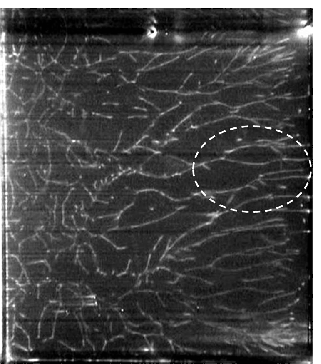}\label{fig:plif_no_shear}}\hfill\\
\subfigure[]{\includegraphics[width=0.225\textwidth]{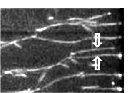}\label{fig:merplif1}}\hfill
\subfigure[]{\includegraphics[width=0.225\textwidth]{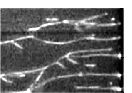}\label{fig:merplif2}}\hfill
\subfigure[]{\includegraphics[width=0.225\textwidth]{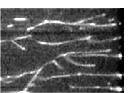}\label{fig:merplif3}}\hfill
\subfigure[]{\includegraphics[width=0.225\textwidth]{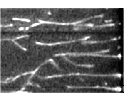}\label{fig:merplif4}}
\caption{Dynamics of plumes observed by
  \cite{puthenveettil05:_plume_rayleig} at $Sc=600$ and $Ra_{w}=2.034
  \times 10^{11}$; (a), The instantaneous planform of the entire
  convection surface of the cell of size $87$mm$\times 98$mm; (b) to
  (e), zoomed views of size $34$mm$\times23$mm of the merging plumes
  in the region showed by the dashed ellipse in (a) at the following
  different times; (b),t=0 s; (c), t=4 s; (d), t=8 s; and (e), t=11
  s.}
  \label{fig:merseqplifnoshear}
\end{figure}
Pairs of line plumes that are near to each other and nearly parallel
were observed to move perpendicular to their length and merge.  Such a
motion could be seen in figures \ref{fig:merair1} to
\ref{fig:merair4}, which shows the zoomed view of the region marked by
the dashed ellipse in the planform in
figure~\ref{fig:air_noshear_planform} at $Ra_{w}=2.54\times10^{8}$ and
$Pr=0.74$.  The lateral merging is obvious from the reducing distance
between the line plumes indicated by the hollow arrows in these images
which are separated by 0.4 s from each other. Similar lateral merging
at around similar $Ra_w=2.65\times10^{8}$ but at a higher $Pr=4.7$ in
water is shown in figures~\ref{fig:merthym1} to \ref{fig:merthym4}.
The merging sequence of two nearby plumes in the concentration driven
experiments at higher $Ra_{w}=2.034\times10^{11}$ and higher $Sc=600$,
than the case of water is seen in figures \ref{fig:merplif1} to
\ref{fig:merplif4}. These merging motions are the predominant dynamics
of these line plumes, especially at lower $Ra$ and higher $Pr$.

The lateral merging of plumes occurred in all the regions with
different strengths of shear in a planform. However, the plumes were
observed to merge faster in the low shear regions than those in the
high shear regions at the same $Ra_w$ and $Pr$. Further, the number of
merging instances were also observed to decrease in high shear areas
compared to the low shear areas at the same $Ra_w$ and $Pr$. In both
the regions with less shear as well as strong shear, the velocities of
merging were observed to increase with increase in $Ra_w$ or decrease
in $Pr$. Due to the presence of large shear in some regions, or due to
the absence of nearby plumes in other regions, the total length of
plumes in the planform does not undergo lateral merging. There is a
fraction, which is a function of $Ra_w$ and $Pr$, of the total plume
length in the planform that undergoes merging. Further, we also notice
that plumes have to be close to each to other to undergo merging,
else they remain stationary or are swept away by the large scale flow.

The merging of nearby plumes were not limited to only when they were
parallel. Often, one end of nearby plumes would be touching each other
with the distance between the plumes increasing as we go away from the
contact point. These plumes also merge in such a condition, often
becoming parallel during the merging process (see figures
\ref{fig:merplif1} to \ref{fig:merplif4} and \ref{fig:merplifshear1}
to \ref{fig:merplifshear4} for plumes touching at P). Often, since
there are more than a pair of plumes that are nearby to each other,
merging of a pair would also result in this pair of plumes moving away
from nearby plumes, we however do not classify this as a separate
motion since the motion is again due to the merging.
\subsection{Initiation of plumes}
\label{sec:initiation-plumes}
\begin{figure}
  \centering
  \subfigure[]{\includegraphics[width=0.24\textwidth]{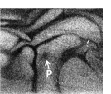}\label{fig:initair1}}
  \subfigure[]{\includegraphics[width=0.24\textwidth]{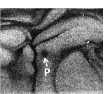}\label{fig:initair2}}
  \subfigure[]{\includegraphics[width=0.24\textwidth]{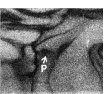}\label{fig:initair3}}
  \subfigure[]{\includegraphics[width=0.24\textwidth]{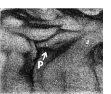}\label{fig:initair4}}
  \caption{Sequence of images showing the dynamics of initiation and
    elongation of line plumes at $Ra_w=2.54\times10^{8}$ and
    $Pr=0.74$. The point P shows the initiation of a line plume. The
    images are of size $171$mm$\times135$mm and are separated by 0.2
    s.}
  \label{fig:initiationair}
\end{figure}
\begin{figure}
  \centering
  \subfigure[]{\includegraphics[width=0.24\textwidth]{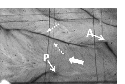}\label{fig:dotplumemerging1}}
  \subfigure[]{\includegraphics[width=0.24\textwidth]{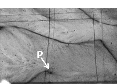}\label{fig:dotplumemerging2}}
  \subfigure[]{\includegraphics[width=0.24\textwidth]{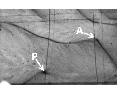}\label{fig:dotplumemerging3}}
  \subfigure[]{\includegraphics[width=0.24\textwidth]{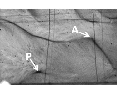}\label{fig:dotplumemerging4}}
  \caption{Dynamics of initiation and elongation of line plumes at
    $Pr=4.7$, and $Ra_{w}=2.65 \times 10^{8}$. Point P shows the
    location of initiation. The dashed arrows show the direction of
    motion in the boundary layers between the plumes. The images are
    of size $98$mm$\times65$mm and are separated by 1.5 s.}
  \label{fig:dotplumemerging}
\end{figure}
The third class of motion that we observe often is the initiation of
new plumes. New plumes are initiated in regions that have become free
of plumes due to either or both of the two class of motions described
in \S~\ref{sec:lateral-motion-line} and \S~\ref{sec:long-moti-line}.
Figure \ref{fig:initiationair} shows the sequence of initiation of a
plume in a region free of plumes at $Ra_w=2.54\times10^{8}$ and
$Pr=0.74$. The initiation occurs as a point burst on the underlying
boundary layer, shown by point P in figure~\ref{fig:initair1}. This
point plume is soon elongated in the direction of the prevalent shear
in the region at which it is initiated (see figures~\ref{fig:initair2}
to \ref{fig:initair4}). The prevalent shear at this point could be due
to the entrainment flow field of a nearby plume, or it could be due to
the external shear by the large scale flow. In the former case the
elongation is often approximately at right angles to the nearby plume;
figure~\ref{fig:initiationair} is an example of such elongation. The
initiation dynamics at a larger $Pr$ of $4.6$ at around the same
$Ra_w$ as in figure~\ref{fig:initiationair} is shown in
figure~\ref{fig:dotplumemerging}. The filled arrow shows the
predominant shear field near point P in these figures as inferred from
the motion of point A in the figures. The shear is mostly due to the
large scale flow since the plumes are seen to be aligned in this
direction, the point plume also elongates in the direction parallel to
the nearby plumes. The initiation dynamics at the location marked by D
in figures~\ref{fig:merplifshear1} to \ref{fig:merplifshear4} at a
much larger $Sc$ of $600$ and at a higher $Ra_w$ of $2.043\times
10^{11}$ shows the influence of both the external shear and the nearby
plumes. The initial elongation of the plume is along the direction in
which the nearby plumes are aligned, parallel to the nearby plumes,
and hence is mostly due to the external shear.  However at later
stages, the orientation of the plume changes so as to join the nearby
plume at an angle, possibly due to the influence of the nearby
plume. By comparing the size of the initiation plume at the three
different $Pr$ in figures~\ref{fig:initiationair},
\ref{fig:dotplumemerging} and~\ref{fig:merseqplifshear} we could see
that the initiation occurs as a smaller, sharper point with increase
in $Pr$. In addition the elongation also appears to be slower with
increase in $Pr$.
\subsection{Miscellaneous motions}\label{sec:misc-moti}
\begin{figure}
  \centering
  \subfigure[]{\includegraphics[width=0.24\textwidth]{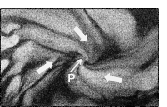}\label{fig:multiplumemergeair1}}
  \subfigure[]{\includegraphics[width=0.24\textwidth]{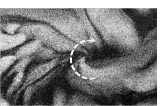}\label{fig:multiplumemergeair2}}
  \subfigure[]{\includegraphics[width=0.24\textwidth]{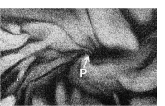}\label{fig:multiplumemergeair3}}
  \subfigure[]{\includegraphics[width=0.24\textwidth]{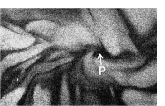}\label{fig:multiplumemergeair4}}
  \caption{(a) to (d), Swirling motion of plumes at $Pr=0.74$, and
    $Ra_{w}=2.54 \times 10^{8}$; (a), the filled arrows show the
    direction of shear while the dashed arrow shows the direction of
    swirling about the point P. The images are of
    size$171$mm$\times112$mm and separated by 0.2 s.}
  \label{fig:multiplumemergeair}
\end{figure}
In addition to the above three major class of motions, there are other
minor motions that occur not so frequently in the planforms. One of
such motions is shown in figure~\ref{fig:multiplumemergeair} which
occurs at the intersection of shear in the opposite directions. In
figure~\ref{fig:multiplumemergeair1} there is a shear directed
downwards from the top left shown by the top filled arrow and shear
upwards from the bottom right, shown by the bottom filled arrow.
Intersection of these two shear streams causes the line plumes to
start swirling about point P in the direction shown by the dashed
arrow. The point P is usually also not stationary, as could be
observed from figure \ref{fig:multiplumemergeair} (a) to (d) where the
location of P has shifted to the right. We find such swirling motion
to occur more at lower $Pr$ at a given $Ra_w$, possibly due to the
stronger shear at the lower viscosity at the lower $Pr$.
\begin{figure}
  \centering
  \subfigure[]{\includegraphics[width=0.24\textwidth]{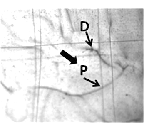}\label{fig:thymlateralshearmerge1}}
  \subfigure[]{\includegraphics[width=0.24\textwidth]{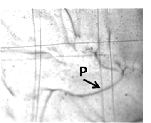}\label{fig:thymlateralshearmerge2}}
  \subfigure[]{\includegraphics[width=0.24\textwidth]{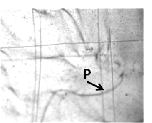}\label{fig:thymlateralshearmerge3}}
  \subfigure[]{\includegraphics[width=0.24\textwidth]{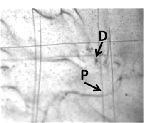}\label{fig:thymlateralshearmerge4}}
  \caption{(a) to (d), Lateral motion of plumes due to shear at
    $Pr=4.7$, and $Ra_{w}=2.65 \times 10^{8}$. The plume P moves
    laterally along the direction of shear, identified by the motion
    of the feature marked by D. The images are of size
    $98$mm$\times79$mm and separated by 2 s.}
  \label{fig:thymlateralshearmerge}
\end{figure}
\begin{figure}
  \centering
  \subfigure[]{\includegraphics[width=0.3\textwidth]{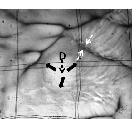}\label{fig:thymdownflow1}}\hfill
  \subfigure[]{\includegraphics[width=0.3\textwidth]{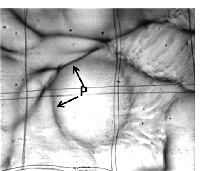}\label{fig:thymdownflow2}}\hfill
  \subfigure[]{\includegraphics[width=0.3\textwidth]{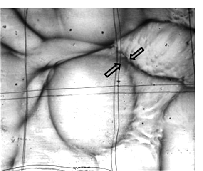}\label{fig:thymdownflow3}}\hfill
  \caption{(a) to (d), Lateral plume motion due to local impingement
    of bulk flow at $Pr=5.3$, and $Ra_{w}=1.31 \times 10^{5}$. The
    filled arrows show the shear due to the downflow at D. The plumes
    P align at the outer periphery of the local shear. The dashed
    arrows show the direction of motion in the boundary layers between
    the plumes. The images are of size $157$mm$\times135$mm and are at
    times 0 s, 8 s and 14 s respectively. }
  \label{fig:thymdownflow}
\end{figure}

Even though plumes usually align along the direction of shear, as
discussed in \S~\ref{sec:long-moti-line}, there are situations where
this does not occur. We sometimes find plumes whose both ends are
connected to other plumes which are aligned along the direction of
shear. Since such a plume is prevented from aligning by the plumes at
its ends, it often moves laterally in the direction of shear without
aligning along the direction of shear.
Figure~\ref{fig:thymlateralshearmerge} shows such an instance where
the plume marked by P moves laterally in the direction of shear
without aligning in this direction.  The filled arrow shows the
direction of shear, which could be inferred by looking at the
successive locations of the feature D in figures
\ref{fig:thymlateralshearmerge1} to \ref{fig:thymlateralshearmerge4}.
A similar situation of lateral plume motion in the presence of shear
is also sometimes seen when the shear is not caused by the large scale
flow, but by the local impingement of flows necessitated by rising
local plume columns. Figure~\ref{fig:thymdownflow} shows such an
instance where there is local impingement of flow at the point D which
drives a shear in the directions shown by the filled arrows, thereby
moving the plume P in a lateral direction. The shear could not be due
to the external large scale flow, since it is not unidirectional as it
is in the case of planforms discussed in \S~\ref{sec:long-moti-line}
and since there is no alignment of any of the plumes around the region
of impingement.

Eventhough we have described the various motions in
\S~\ref{sec:long-moti-line} to \ref{sec:misc-moti} for the sake of
completeness, all these motions are not equally important at any
instant. One could detect the predominant motions by qualitatively
estimating the fraction of the area of the planform at any instant
that exhibit these motions. Such an evaluation shows that the most
predominant motion is merging, followed by shear and then initiation,
the miscellaneous motions occuring rarely.
\section{Measurement of plume dynamics}\label{sec:meas-plume-dynam}
As mentioned in \S~\ref{sec:misc-moti}, the predominant motion of
plumes in terms of area affected, are merging, shear and initiation.
We undertake the following five measurements to quantify these
dynamics described qualitatively in \S~\ref{sec:long-moti-line},
\S~\ref{sec:lateral-motion-line} and \S~\ref{sec:initiation-plumes}.
\begin{enumerate}
\item The velocities along the line plumes ($V_{sh}$).
\item The spacing ($\lambda$) between two line plumes that are nearly
  parallel as a function of time.
\item The fraction of the total length of plumes over which there is
  substantial longitudinal motion ($L_{ps}/L_p$).
\item The fraction of the total length of plumes in a planform that
  merge ($L_{pm}/L_p$).
\item The time of initiation of point plumes $t^*$.
\end{enumerate}  
All these measurements are made on top view images, similar to that
shown in figures \ref{fig:merseqairnoshear} to
\ref{fig:merseqplifshear}
\subsection{Velocities along plumes}
\label{sec:veloc-along-plum}
The velocities along the length of the plumes were calculated by
measuring the speed at which some feature of the plume moves along the
plume.  For example, we estimate the velocity along the plumes in
figure \ref{fig:merthymshear1} to \ref{fig:merthymshear4}, by
measuring the displacement of the dot in the plume structure, marked
by P, along the plume length in successive frames of the merging
sequence. Since the plumes align easily along the shear direction, we
expect the velocity estimated in this way to be indicative of the
strength of shear caused by the large scale flow.  The feature is
tracked over a time period of $1.5$ to $13 s$ which is much lower than
the time scale over which the large scale flow changes. The feature is
tracked over a distance of the order of less than one tenth of the
tank width so that effects of spatial variations in shear, expected to
be substantial only over half the tank width, could be expected to be
negligible in these local measurements.  Hence, we use a linear fit
between the position of the plume feature with time to calculate the
derivative with time to estimate a constant local velocity along the
plume at a given $Ra_{w}$ and $Pr$ at any location.
\subsection{Plume spacing as a function of
  time}\label{sec:merging-velocities}
The merging velocities of plumes were calculated by measuring the
position of two merging line plumes from successive images of a
merging sequence. A merging sequence of a pair of line plumes was
first identified by viewing a video recording of the time evolution of
the whole planform of plume structure. Figure \ref{fig:merair1}, and
other similar planforms at different $Pr$, show plumes identified in
this way within a dashed ellipse. The co-ordinates of two points- one
each on each facing edge of both the plumes in the identified plume
pair - that could be connected by a line approximately perpendicular
to these line plumes, are captured from mouse clicks by a program. The
same process is repeated in each successive image of the merging
sequence.  The program then calculates the distance between these
points in each frame of the merging sequence to obtain the plume
spacing $\lambda$ as a function of time $t$, the time being calculated
from the frame rate of the recording. The maximum error involved in
the measurement of $\lambda$ were of the order of $1\%$ as estimated
from the pixel size of the images. These measurements were then
repeated for different merging sequences at different times and
locations in the video of the time evolution of the planform for each
$Ra_w$.
\subsection{Merging and sheared plume
  lengths}\label{sec:length-merg-plum}
\begin{figure}
  \centering
 \includegraphics[width=0.4\textwidth]{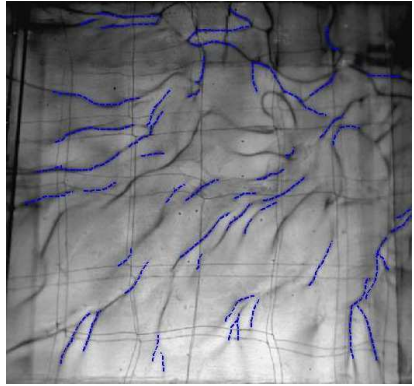}
 \caption{Measurement of length of merging plumes ($L_{Pm}$) at
   $Ra_{w}=2.83\times10^{7}$ and $Pr=5.08$. The dashed lines show the
   parts of the total plume structure that undergo merging}.
  \label{fig:linemark}
\end{figure}
Information on the fraction of the total plume length that is merging
at any $Ra_w$ is needed to understand the importance of the merging
motion in the total dynamics of the plumes near the plate. We
determine this fraction in the following way. A short video clip, of
time duration of few seconds, is made from the complete video of the
time evolution of the planforms. This video clip is played
continuously at a much higher speed than the actual recording speed so
that merging parts of the plumes could be easily identified.  Along
with this movie clip, the first planform image from this movie video
clip is viewed. Using a program that marks lines on images and
estimates the total length of such
lines~\cite*[][]{puthenveettil11:_lengt} the regions of plumes that
are merging in the movie clip are marked with short linear segments in
the planform image.  Figure \ref{fig:linemark} shows an example of
such a marked planform, where the blue lines are the parts of the
plumes that show substantial lateral motion. The program then adds up
the total length of the marked lines to give an estimate of the length
of plumes that are merging at any $Ra_w$. A similar procedure is also
used to estimate the fraction of the total length of plumes in a
planform at a specific $Ra_w$ and $Pr$ that have substantial
longitudinal motion along the plumes.
\subsection{Time of initiation of plumes}
\label{sec:merg-shear-plume}
As discussed in \S~\ref{sec:initiation-plumes}, line plumes are often
initiated as points which then elongate in the prevalent direction of
shear at the initiation point. We measure the time for the initiation
of these plumes as points, with the intention to understand the
process by which these plumes are generated.  Movies of the evolution
of the planforms are repeatedly played while focussing attention in a
specific region of the plan form so as to spot a location at which a
plume is initiated as a point, similar to that shown in figures
\ref{fig:initiationair}, ~\ref{fig:dotplumemerging} and
\ref{fig:merseqplifshear}. Once a location of initiation is
identified, the movie is stopped at that frame and the movie is played
in reverse while examining the frames so as to find another frame in
which there is no plume at the location of the initiation.  The time
between these two frames is taken as the time of initiation of a plume
$t^*$ at that location. This procedure is repeated by identifying
similar initiation of plumes at different locations on the planforms
for a given $Ra_{w}$ and $Pr$. The mean time of initiation of plumes
$\overline{t^{*}}$ at a given $Ra_w$ and $Pr$ is determined from such
multiple measurements.
\section{Dynamics of line plumes}\label{sec:dynamics-line-plumes}
We now quantify the three main dynamics that we described
qualitatively in \S~\ref{sec:long-moti-line},
\S~\ref{sec:lateral-motion-line} and \S~\ref{sec:initiation-plumes}.
We focus more on quantifying the dynamics of merging of plumes since
far less is known on this type motion compared to the other two.
\subsection{Longitudinal motion of plumes}
\label{sec:shear-along-plumes}
\begin{figure}
  \centering
  \subfigure[]{\includegraphics[width=0.485\textwidth]{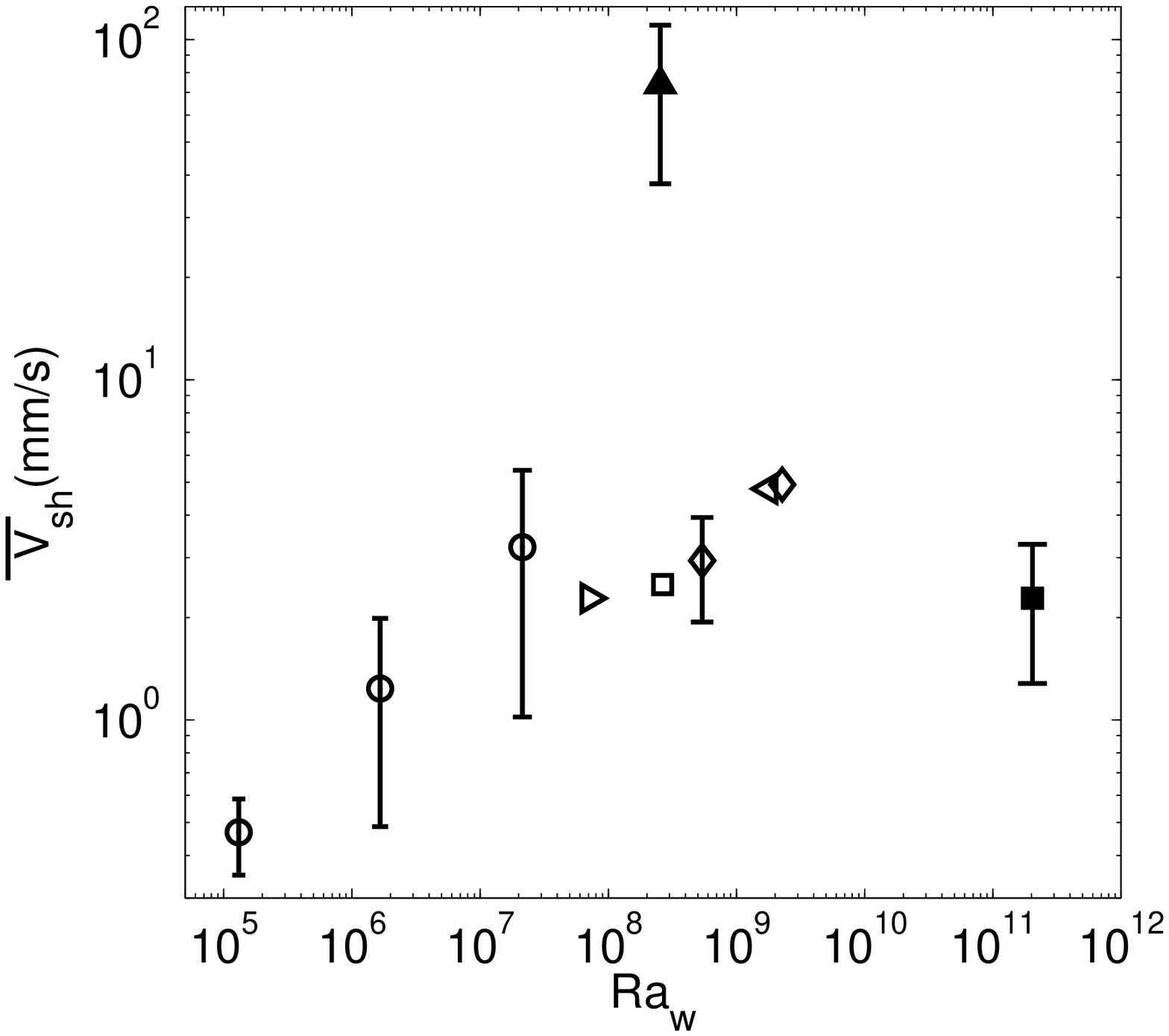}\label{fig:meanshear}}\hfill
  \subfigure[]{\includegraphics[width=0.485\textwidth]{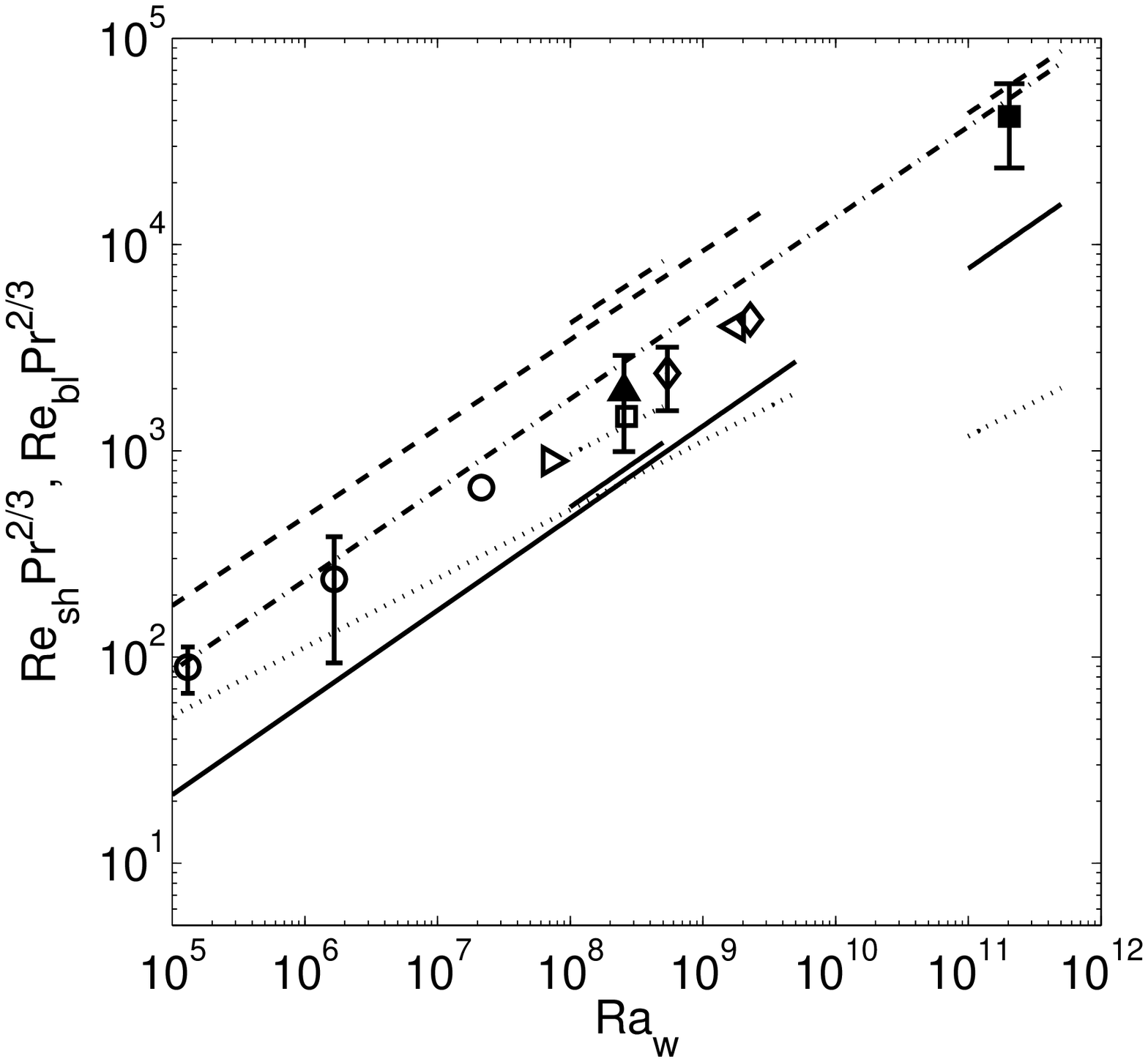}\label{fig:comp_shear_BL}}
  \caption{(a), Variation of the mean longitudinal velocity
    $\overline{V_{sh}}$ with $Ra_{w}$ for the three $Pr$. The open
    symbols show $\overline{V_{sh}}$ for the convection experiments in
    water for $Pr=3.6$ to $5.3$ for the following layer heights;
    $\circ,\, H= 50$mm; $\rhd,\, H= 100$mm; $\Box,\, H=150$mm;
    $\bigtriangledown,\,H=200$mm and $\diamond,\, H=210$mm.
    $\blacktriangle$ represent experiments at $Pr=0.74$ and $H=500$mm
    while $\blacksquare$ represent experiments at $Sc=602$ and
    $H=230$mm; (b), Variation of the Reynolds number based on the mean
    longitudinal velocity of motion along the plumes with
    $Ra_{w}$. $-\cdot-\cdot
    0.55Ra_w^{4/9}Pr^{-2/3}$\cite*[][]{puthenveettil05:_plume_rayleig},
    $- - - 1.09
    Ra^{0.43}Pr^{-0.76}$\cite*[][]{xia02:_heat_prand_rayleig},
    \textemdash~
    $0.102Ra^{0.447}Pr^{-2/3}$\cite*[][]{ahlers09:_heat_rayleig_benar_convec}
    and $\cdot\cdot\cdot\cdot 0.19 Ra_w^{1/3}Pr^{-0.98}$
    (\ref{eq:rebl}) shown for each of the $Pr$ separately.}
  \label{fig:shearonplumes}
\end{figure}
Figure~\ref{fig:meanshear} shows the variation of the mean velocity of
longitudinal motion along the plumes $\overline{V_{sh}}$, measured as
discussed in \S~\ref{sec:veloc-along-plum}, at various $Ra_w$ and
$Pr$. $\overline{V_{sh}}$ was calculated from measurements at about
five locations at various times from movies of plume motion at a given
$Ra_w$ and $Pr$. The vertical bars in the figure show the range of
variation of the longitudinal velocity at each $Ra_w$ and $Pr$. The
mean longitudinal velocities increase with increase in $Ra_w$ and
decrease in $Pr$. At the lower $Ra_w$ in water, the velocities are
small ($\sim 0.4$mm/s) but become appreciable with increase in $Ra_w$
and is quite substantial at about $10$cm/s in air at an $Ra_w$ of
$10^8$. There also seems to be a trend of increasing range of
longitudinal velocities with increasing $Ra_w$ at the same $Pr$. Since
these behaviours are also exhibited by the large scale flow in
turbulent convection we expect the longitudinal motion of plumes to be
caused by the shear of the large scale flow; we now verify this by
comparing $\overline{V_{sh}}$ with the large scale flow velocities in
turbulent convection.

Figure~\ref{fig:comp_shear_BL} shows the expressions for the large
scale flow strength, in terms of the corresponding Reynolds number,
proposed by \cite{puthenveettil05:_plume_rayleig},
\cite{xia02:_heat_prand_rayleig} and Grossman \& Lohse (2009), along
with the variation of $Re_{sh}=\overline{V_{sh}}H/\nu$, the Reynolds
number based on the mean longitudinal velocity on the plumes. The
magnitude and scaling of $Re_{sh}$ is almost the same as that of the
large scale flow proposed by \cite{puthenveettil05:_plume_rayleig},
i.e.,
\begin{equation}
  \label{eq:resh}
  Re_{sh}=0.55Ra_w^{4/9}Pr^{-2/3}.
\end{equation}
In addition, since there is no other obvious force, other than that
caused by the shear of the large sale flow, that would be needed to
balance the viscous resistance for the longitudinal motion along the
plumes, it would be plausible to conclude that the longitudinal motion
along the plumes is caused by the shear of the large scale flow on
these plumes. 

This shear on the plumes also act on the boundary layers in between
the plumes. \cite{puthenveettil11:_lengt}, without considering the
presence of such a shear, have shown that the observed length of these
lines plumes in any area of the plate could be explained by the
phenomenology of natural convection boundary layers becoming unstable
at $Ra_\delta\sim 1000$; a similar phenomenology also explains the
scaling of mean spacing between the
plumes~\cite*[][]{theerthan98:_rayleig,puthenveettil05:_plume_rayleig}.
Judging from the match obtained by these theories with the experiments
on the mean spacing and the total length of plumes, the effect of
shear on the assumed phenomenology, viz. on the stability of these
local boundary layers and on the flux scaling, is mostly
small. However, as found out recently by
\cite{shi12:_bound_rayleig_benar}, the velocity profiles near the
plate in experiments do not match the profiles from these theories
that assume natural convection boundary layers. Note that such a
comparison was not done on the spatially averaged profiles arising out
of two natural convection boundary layers giving rise to a plume,
which is necessary, as shown by \cite{theerthan98:_rayleig}. Still, as
first pointed out by \cite{puthenveettil05:_plume_rayleig}, and
recently proposed by \cite{shi12:_bound_rayleig_benar}, with
increasing $Ra_w$ the local boundary layers between the plumes will be
affected by the external shear of the large scale flow.
\begin{figure}
  \centering
  \includegraphics[width=0.6\textwidth]{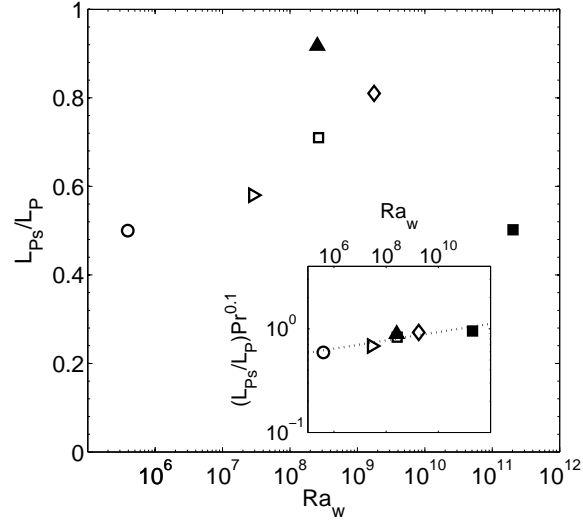}
  \caption{Variation of the fraction of the total plume length that
    have longitudinal motion along them ($L_{ps}/L_p$) with $Ra_w$ and
    $Pr$. The open symbols indicate thermal convection experiments in
    water for $Pr=3.6$ to $5.3$ for the following layer heights;
    $\circ,\, H= 50$mm; $\rhd,\, H= 100$mm; $\Box,\, H=150$mm; and
    $\diamond,\, H=210$mm. $\blacktriangle$ represents an experiment
    at $Pr=0.74$ and $H=500$mm while $\blacksquare$ represents an
    experiment at $Sc=602$ and $H=230$mm. The inset shows the scaling
    of $L_{ps}/L_p$. $.... 0.37Ra_w^{0.04}Pr^{-0.1}$ }
  \label{fig:lpshbylp}
\end{figure}

Such an inference could be formed by comparing the magnitudes of the
horizontal velocities inside the local boundary layers feeding the
plumes and the large scale flow
velocities. Figure~\ref{fig:comp_shear_BL} shows the variation of
$Re_{bl}=u_{bl}H/\nu$, the Reynolds number in terms of $u_{bl}$, the
horizontal characteristic velocity inside the natural convection
boundary layers feeding the plume from each side at its base, as given
by (A12) in \cite{puthenveettil11:_lengt}.
\begin{equation}
  \label{eq:rebl}
  Re_{bl}= \left(\frac{c_1}{2}\right)^{1/5} Ra_w^{1/3}Pr^{\frac{n_1}{5}-1},
\end{equation}
as obtained from (A12) and (A11) of \cite{puthenveettil11:_lengt}, by
using (\ref{eq:zw}) and (\ref{meanpluspace}), where, $C_1=47.5$ and
$n_1=0.1$.  $Re_{bl}$ is an order of magnitude smaller than $Re_{sh}$,
the difference in magnitudes between the Reynolds numbers increasing
with increasing $Ra_w$. Eventhough the value of external shear given
by (\ref{eq:resh}) is at a much higher location that the height at
which the natural convection boundary layers turn into a plume (see
\cite{puthenveettil11:_lengt}), at larger $Ra_w$, the large scale flow
will affect the flow inside the local natural convection boundary
layers to turn them into mixed convection boundary layers.

Such an effect of the external shear on the boundary layers could be
detected, eventhough qualitatively, from our visualisations. The
dashed arrows in figure~\ref{fig:thymdownflow1} shows the motion in
the boundary layers in between the plumes in a region where there is
no substantial shear. The dashed arrows are drawn parallel to the dye
streaks seen in the figure; the streaks are formed when the dye layer
formed on the plate gets drawn in the predominant motion direction.
The motion is approximately perpendicular to the plumes, as would be
expected if the boundary layers were natural convection boundary
layers feeding the plume at its base. However in
figure~\ref{fig:dotplumemerging1}, which due to the higher $Ra_w$ has
more external shear, the motion in the boundary layers in between the
plumes is no longer perpendicular to the plumes. The dashed arrows in
the figure show that the motion in the boundary layers is inclined in
the direction of the external shear, shown by the thick filled white
arrow; one could expect these boundary layers to be of mixed
convection. At any instant in the planforms at any $Ra_w$ and $Pr$,
there are regions with substantial shear along the plumes as well as
without it. Hence, the boundary layers feeding the plumes would show a
varying nature between pure natural convection type and mixed
convection type at various locations, based on the relative magnitude
of the external shear and the characteristic velocity of natural
convection boundary layers at each location; the former being given by
(\ref{eq:resh}), while the latter being given by (\ref{eq:rebl}).

It is hence important to get an idea of the extent of the bottom plate
area affected by the external shear at different $Ra_w$ and
$Pr$. Figure \ref{fig:lpshbylp} shows the variation of $L_{ps}/L_p$,
the fraction of the length of plumes that have noticeable shear along
the length of the plumes, for various $Ra_w$ and three $Pr$;
$L_{ps}/L_p$ were measured as discussed in
\S~\ref{sec:length-merg-plum}. The region affected by shear increases
with increase in $Ra_w$ and with decrease in $Pr$, both these changes
also increase the strength of the large scale flow as per
(\ref{eq:resh}). The inset in the figure shows that
\begin{equation}
  \label{eq:lpslp}
  \frac{L_{ps}}{L_p}= 0.37Ra_w^{0.04}Pr^{-0.1}.
\end{equation}
Interestingly, in \S~\ref{sec:fract-merg-length}, we find an inverse
dependence on $Ra_w$ and $Pr$ of that in (\ref{eq:lpslp}) for the
fraction of plume lengths that is merging.
\subsection{Initiation of plumes}
\label{sec:initiation-plumes-1}
\begin{figure}
  \centering
  \includegraphics[width=0.6\textwidth]{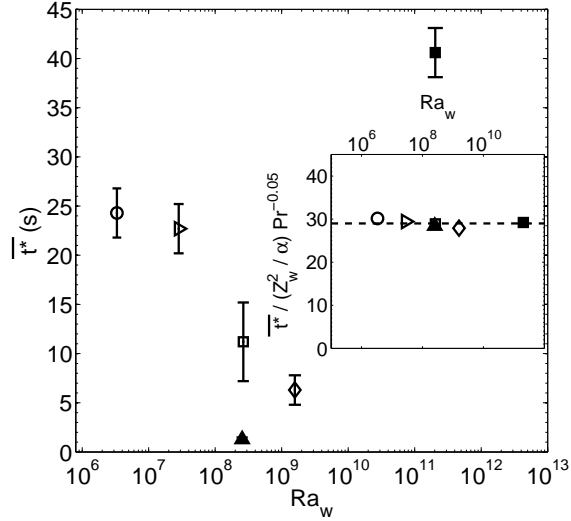}
  \caption{Variation of the mean time for initiation of plumes as
    points with $Ra_{w}$ at the three $Pr$. The open symbols indicate
    thermal convection experiments in water for $Pr=3.6$ to $5.3$ for
    the following layer heights; $\circ,\, H= 50$mm; $\rhd,\, H=
    100$mm; $\Box,\, H=150$mm; and $\diamond,\, H=210$mm.
    $\blacktriangle$ represents the mean from experiments at $Pr=0.74$
    and $H=500$mm while the $\blacksquare$ represents the mean from
    experiments at $Sc=602$ and $H=230$mm. The error bar gives the
    range of time of initiation of plumes over the planform for a
    given $Ra_{w}$. The inset shows that the dimensionless time of
    initiation of plumes is a constant for a given $Pr$.}
  \label{fig:timeplumeinit}
\end{figure}

It was observed in \S~\ref{sec:initiation-plumes} that plumes usually
initiate as points in regions that are free of line plumes, the time
for such initiation was measured as discussed in
\ref{sec:merg-shear-plume}. Figure \ref{fig:timeplumeinit} shows the
variation of the mean time of initiation of point plumes at various $Ra_w$
and $Pr$. Plumes initiate faster at larger $Ra_w$ and at lower
$Pr$. As per \cite{how}, initiation of plumes occur by periodic
eruption of the boundary layer that grows by a diffusive process as
$\delta(t)=\sqrt{\pi\alpha t}$. The boundary layer erupts at a time
$t^*$ at which the Rayleigh number based on the boundary layer
thickness $Ra_\delta =g\beta \Delta T_w \delta^3/\nu\alpha$ becomes
approximately $1000$. Using the value of $\delta(t^*)$ in the
expression for $Ra_\delta$ and using (\ref{eq:zw}) to rewrite in terms
of $Z_w$, we obtain
\begin{equation}
  \label{eq:t*}
  t^*\sim\frac{100}{\pi}\frac{Z_w^2}{\alpha},
\end{equation}
with an unspecified $Pr$ dependence, since the stability condition
$Ra_\delta \sim 1000$ has an unknown $Pr$
dependence~\cite*[][]{puthenveettil11:_lengt}. The time of initiation
of plumes is proportional to $Z_w^2/\alpha$, the diffusive time scale
near the plate. The inset in figure~\ref{fig:timeplumeinit} shows that
the measured mean times of initiation of the point plumes scale as
\begin{equation}
  \label{eq:t*exp}
  \overline{t^*}=29 \frac{Z_w^2}{\alpha} Pr^{0.05},
\end{equation}
quite close to (\ref{eq:t*}).
\subsection{Merging of plumes}
\label{sec:merging-plumes}
\subsubsection{Fraction of merging length}
\label{sec:fract-merg-length}
\begin{figure}
  \centering
  \includegraphics[width=0.6\textwidth]{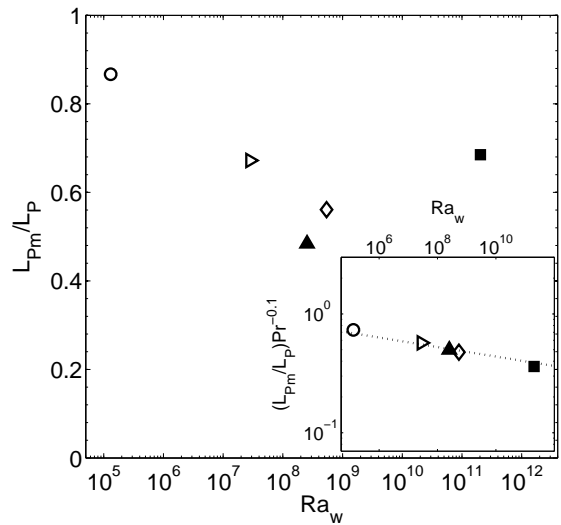}
  \caption{Variation of the ratio of the length of plumes that are
    merging in a planform to the total length of plumes in the
    planform ($L_{pm}/L_p$) with $Ra_{w}$.  The open symbols indicate
    thermal convection experiments in water for $Pr=3.6$ to $5.3$ for
    the following layer heights; $\circ,\, H= 50$mm; $\rhd,\, H=
    100$mm; $\Box,\, H=150$mm; and $\diamond,\, H=210$mm.
    $\blacktriangle,\, Pr=0.73$; $\blacksquare,\, Sc= 600$. The inset
    shows the scaling of $L_{pm}/L_{p}$ as per (\ref{eq:lpmlp}).}
  \label{fig:lpmbylp}
\end{figure} 
As discussed in \S~\ref{sec:lateral-motion-line}, the complete length of
a plume line is not in merging motion at any instant. Parts of the
plume line which are not in merging motion often occur in regions of
larger shear, since shear reduces the merging velocity.  In such
regions there will be substantial motion along the length of the plume
line. In addition, there are also regions in which plumes are not
merging, even when there is no substantial shear in that region, since
there is no line plume nearby to any plume for merging to occur. The
occurrence of such different types of regions on a planform at any
instant necessitates the need to study the role of the merging motion
in the complete dynamics of plumes near the plate.

The ratio $L_{pm}/L_p$, where $L_{pm}$ is the total length of plumes
that are merging in a planform at an instant, measured as discussed in
\S~\ref{sec:length-merg-plum}, is an indicator of the importance of
the merging dynamics in the overall dynamics near the
plate. Figure~\ref{fig:lpmbylp} shows the variation of $L_{pm}/L_p$
with $Ra_w$ measured in the three fluids.  Merging is the predominant
dynamics at the lowest $Ra_w$ in water, with almost 90\% of the length
of the plumes undergoing a merging motion at this $Ra_w$. However with
increase in $Ra_w$, the fraction of length of plumes that display
merging motion in water decreases to about 55\% at $Ra_w\approx
10^9$. Similarly, it is clear that the merging length fraction
decreases with decrease in $Pr$ by comparing the data of air at
$Ra_w\approx 2.5\times 10^8$ with that of water. The inset of
figure~\ref{fig:lpmbylp} shows that
\begin{equation}
  \label{eq:lpmlp}
  \frac{L_{pm}}{L_p}= 1.1Ra_w^{-0.04}Pr^{0.1}.  
\end{equation}
We expect this behaviour of $L_{pm}/L_p$ to be due to the increase in
shear on the plumes, associated with the increase in $Ra_w$ and
decrease in $Pr$. As discussed earlier, the effect of this increased
shear is to decrease the merging velocities. Note that
(\ref{eq:lpmlp}) and (\ref{eq:lpslp}) have an exact inverse dependence
on $Ra_w$ and $Pr$. Such a relationship implies that
\begin{equation}
  \label{eq:lpslpm}
  \frac{L_{ps}L_{pm}}{L_p^2}=0.4,
\end{equation}
a dimensionless invariant number for all the fluids in thermal
convection.
\subsubsection{Variation of merging velocities during the merging
  process }
\label{sec:vari-merg-veloc}
\begin{figure}
  \centering
  \subfigure[]{\includegraphics[width=0.33\textwidth]{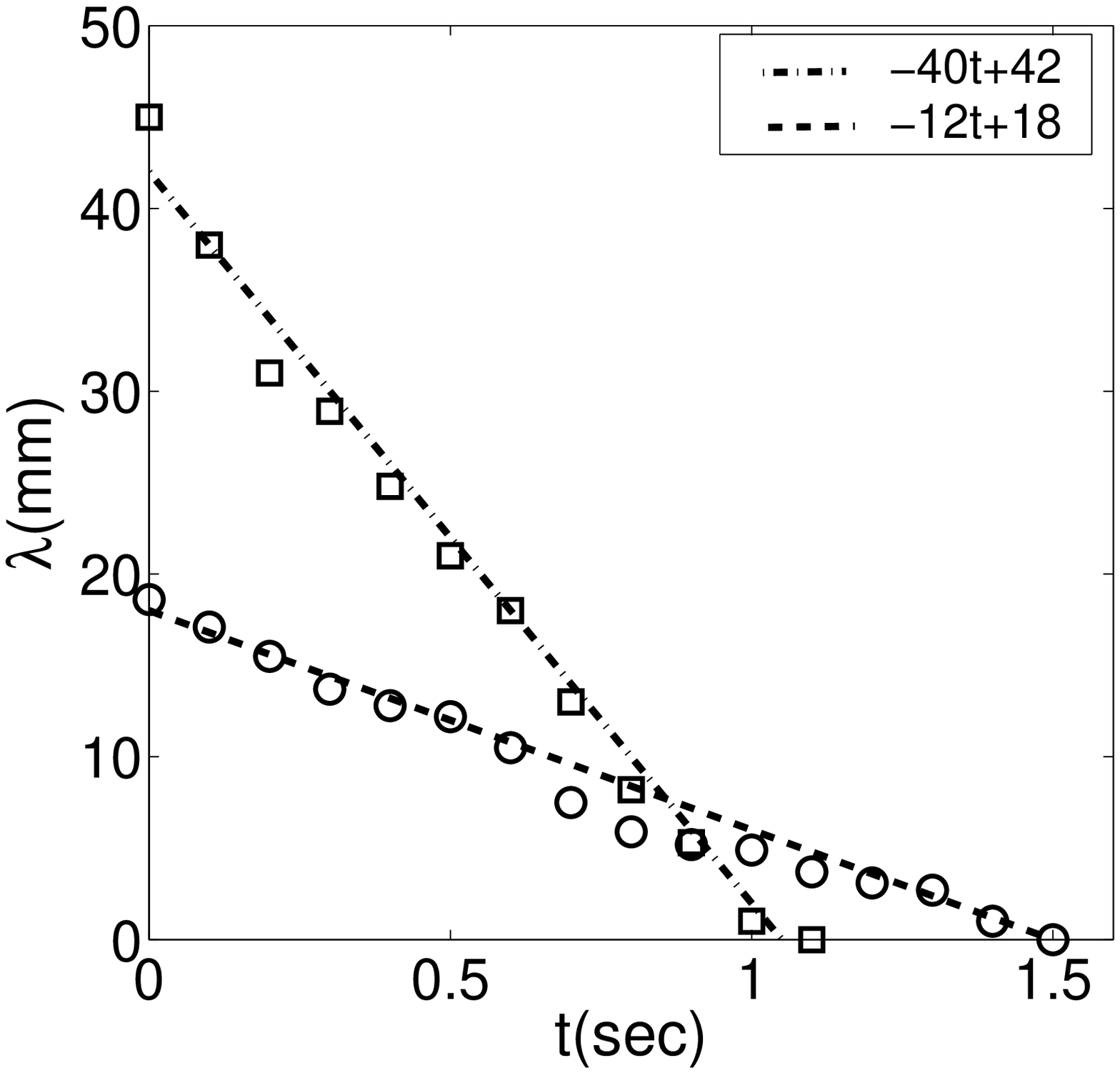}\label{fig:fitair}}\hfill
  \subfigure[]{\includegraphics[width=0.33\textwidth]{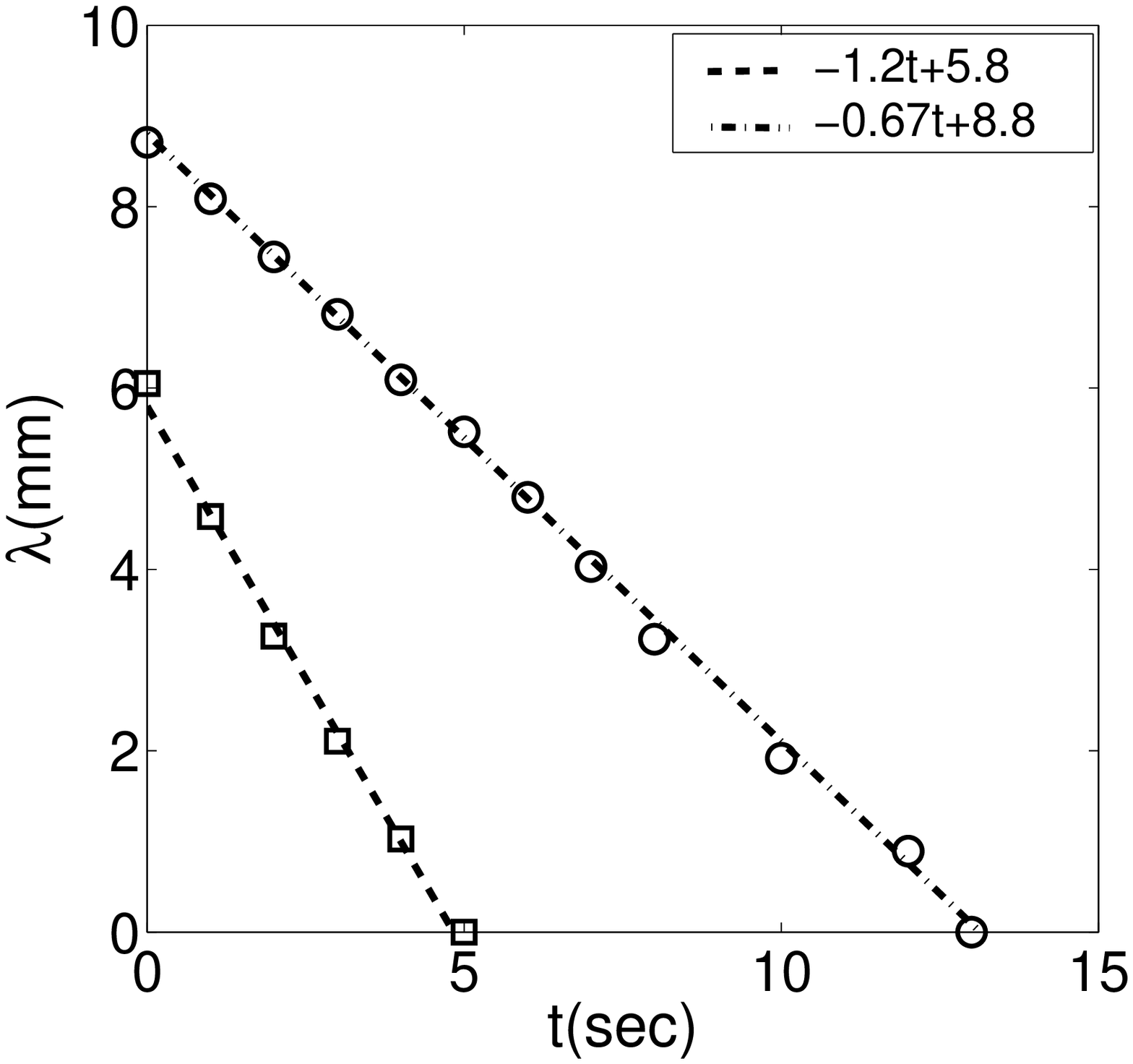}\label{fig:fitthym}}\hfill
  \subfigure[]{\includegraphics[width=0.33\textwidth]{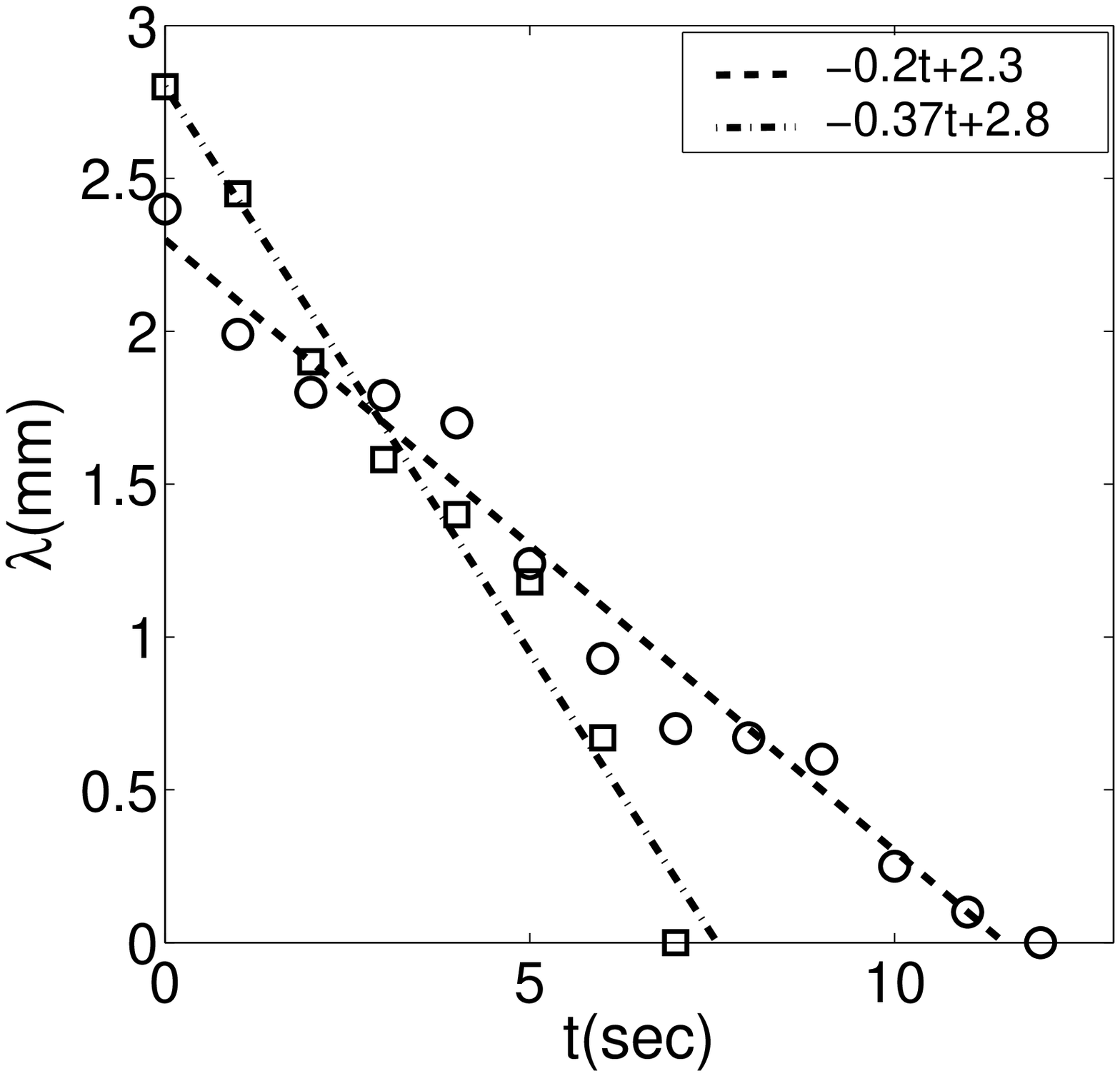} \label{fig:fitplif}}
  \caption{Measured variation of plume spacings~($\lambda$) as a
    function of time ($t$) in regions with shear ($\circ$) and without
    shear ($\Box$) for; (a), $Pr=0.74$ at $Ra_{w}=2.54 \times 10^{8}$;
    (b), $Pr=4.7$ at $Ra_{w}=2.65 \times 10^{8}$ and; (c), $Pr=602$ at
    $Ra_{w}=2.034 \times 10^{11}$.}
\label{fig:fit}
 \end{figure}
 Figure~\ref{fig:fitair} shows the variation of $\lambda$ with time
 for air, obtained by measurements as described in
 \S~\ref{sec:merging-velocities} at the two different locations
 shown in figures~\ref{fig:air_noshear_planform}
 and~\ref{fig:merge_shear_full_planform} at $Ra_w=2.54\times 10^8$.
 The results of similar measurements corresponding to the locations in
 figures~\ref{fig:water_merge_no_shear} and
 ~\ref{fig:merge_shear_water_full} for $Pr=4.7$ and corresponding to
 figures~\ref{fig:plif_no_shear} and ~\ref{fig:merge_shear_plif_full} for
 $Sc=600$ are shown in figures \ref{fig:fitthym} and \ref{fig:fitplif}
 respectively.  It could be noticed that the variation of $\lambda$ in
 all these plots is approximately linear with time.  Such linear
 behaviour of $\lambda$ with $t$ were obtained for all the
 measurements at different $Ra_w$ for all the three $Pr$ at all the
 locations of a planform. The maximum
 variation of the data from linearity was of the order of $5\%$ from
 all the curve fits. Due to such a linear variation of $\lambda$ with
 time at all $Ra_w$ and $Pr$, we estimate the merging velocities of
 plumes by calculating the gradient of a linear curve fit through the
 measured $\lambda$ vs $t$.
 
 A linear variation of $\lambda$ with $t$ would imply that the plumes
 merge with a constant velocity during their merging period,
 eventhough the value of this constant velocity is itself different at
 different locations at the same $Ra_{w}$, as shown by
 figure~\ref{fig:fit}. In addition, these merging velocities are
 strong functions of $Ra_{w}$ and $Pr$, as is obvious from
 figures~\ref{fig:fitair} to \ref{fig:fitplif}.
\begin{figure}
   \centering
   \includegraphics[width=0.6\textwidth]{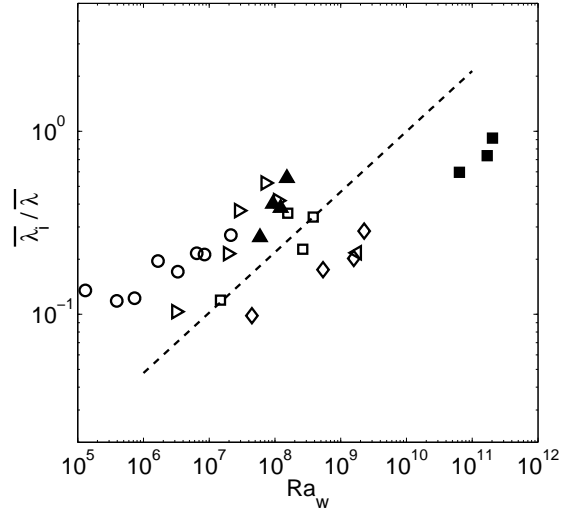}
   \caption{Variation with Rayleigh number of the initial plume
     spacing for merging sequences, averaged over their range at a
     given $Ra_{w}$ and normalised by the mean plume spacing at the
     same $Ra_{w}$. The open symbols indicate thermal convection
     experiments in water for $Pr=3.6$ to $5.3$ for the following
     layer heights; $\circ,\, H= 50$mm; $\rhd,\, H= 100$mm; $\Box,\,
     H=150$mm; $\bigtriangledown,\,H=200$mm and; $\diamond,\,
     H=210$mm.  $\blacktriangle,\, Pr=0.73, H=500$mm; $\blacksquare,\,
     Sc= 600, H=230$mm. $---, 0.0005 Ra_w^{1/3}.$}
   \label{fig:lambdaibylambdac}
 \end{figure}

 It is possible that such a linear variation of $\lambda$ with $t$,
 resulting in a constant merging velocity $V_{m}= \lambda'/2$ where
 $\lambda'= d\lambda/dt$, is because we detect two line plumes as
 merging only when they come close together, so that a nonlinear
 variation appears linear due to the short range measurement. To
 verify whether this is the case for our measurements, we plot the
 variation of $\overline{\lambda_i}/\overline{\lambda}$ for all the
 measurements of $\lambda$ in figure~\ref{fig:lambdaibylambdac}. Here,
 $\overline{\lambda_i}$ is the first measurement of the plume spacing
 measured in each plume merging sequence, averaged over all such
 measurements at the same $Ra_w$. $\overline{\lambda}$ is the critical
 plume spacing given by \cite{theerthan98:_rayleig} and
 \cite{puthenveettil05:_plume_rayleig} as,
\begin{equation}
\overline{\lambda}= C_1Pr^{n_1}Z_w,
\label{meanpluspace}
\end{equation}
where $C_1=47.5$ and $n_1=0.1$. The figure shows that all the
measurements of merging plumes were conducted on plumes that were
separated by an initial spacing smaller than $\overline{\lambda}$.
$\overline{\lambda_i}/\overline{\lambda}$ scale approximately as
$Ra_w^{1/3}$ mainly because all the $\overline{\lambda_i}$ are
approximately equal to 5 mm, except for $Pr=0.74$ where it is about
2.5 cm, while $\overline{\lambda}$ decrease as $Ra_{w}^{-1/3}$ as per
(\ref{meanpluspace}).  As we discussed in
\S~\ref{sec:lateral-motion-line}, only plumes closer than the critical
spacing seems to merge, others fade away or are swept along their
length by the shear. It appears that in our measurements appreciable
merging velocity is detected only when plumes come close to a distance
of the order of $\overline{\lambda}$. However, the constant merging
velocity during a merging cycle is unlikely to be due to the short
distance over which the merging is measured since, as we show later in
\S~\ref{sec:relat-merg-veloc}, there is a strong physical reason for
it to be so.
 \subsubsection{Statistics of merging velocities}
\label{sec:stat-merg-veloc}
\begin{figure}
  \centering
  \includegraphics[width=0.7\textwidth]{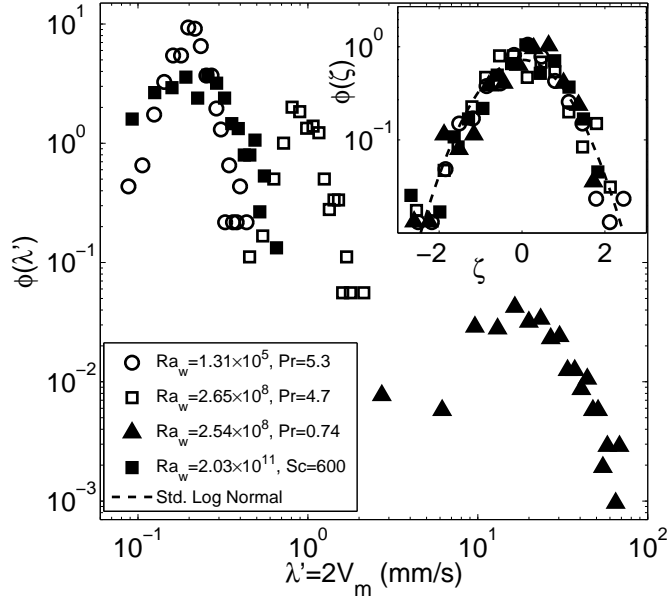}
  \caption{Distribution of merging velocity of plumes ($V_{m}$) for
    different $Ra_{w}$ and $Pr$. The inset shows the data in the main
    figure plotted as the pdf of the logarithm of the dimensionless
    merging velocities in their standardised form $\zeta=
    (\ln(V_{m}/\overline{V_{m}})-\overline{\ln(V_{m}/\overline{V_{m}})})/\sigma\ln(V_{m}/\overline{V_{m}})$. }
\label{rawpdf}
\end{figure}
Figure \ref{fig:fit} shows that plumes in low shear regions merged
faster than those in the shear dominated regions. Since the shear due
to the large scale flow is also a function of space
\cite*[][]{puthenveettil05:_plume_rayleig,PhysRevE.58.5816}, the
merging velocities ($V_{m}$) could be expected to be a function of the
location of the merging plumes. In addition, the presence of nearby
line plumes could also affect the merging velocity of a pair of
plumes. Hence at any instant, $V_{m}$ in a planform are distributed
over a range of values. On the other hand, at any location, for a
given $Ra_w$ and $Pr$, in addition to the fact that plumes are
initiated at different times, the $V_{m}$ that occur in different
merging sequences could also be a function of time since the shear at
a location is also not constant due to the azimuthal rotations and
reversals of the large scale flow
(\cite{ahlers09:_heat_rayleig_benar_convec}). Hence, at any $Ra_{w}$
and $Pr$, the merging velocities $V_{m}$ are functions of space and
time.

We estimate this distribution of $V_{m}$ at any $Ra_w$ and $Pr$ by
measuring the merging velocities from various merging sequences at
different locations and times for that $Ra_w$ and $Pr$.  At each
$Ra_{w}$, about 30 plume merging sequences on the planforms were
identified at different times from regions with and without
shear. From these sequences, a total of about 280 plume merging
velocities were measured.  To increase the number of samples of
$V_{m}$, the velocities during any merging sequence were calculated by
estimating $d\lambda/dt$ from successive values of $\lambda$ in time,
and not from a curve fit as in \S~\ref{sec:merging-velocities}.

Figure \ref{rawpdf} shows the probability density function (pdf) of
$V_{m}$ obtained from these measurements at four different $Ra_{w}$
for the three $Pr$. Comparing the $Pr=4.7$ and $0.74$ curves in the
figure ($\square$ and $\blacktriangle$), which are approximately at
the same $Ra_w$, it is clear that a decrease in $Pr$ results in a
higher mean merging velocity ($\overline{V_{m}}$). Plumes merged
faster at a lower $Pr$ at the same $Ra_w$. Since viscous effects will
be lower in a lower $Pr$ fluid, the resistance to merging motion will
be lower, the mean merging velocities could hence be expected to be
larger.  However, lower $Pr$ also results in larger shear velocities
due to the stronger large scale
flow~\cite*[][]{ahlers09:_heat_rayleig_benar_convec}; as we saw
earlier, this shear slows down the merging. Such a slowing of merging
due to the increase in the strength of the large scale flow do not
seems to be the dominating effect compared to the possible increase in
merging velocities due to the reduction in viscous resistance to
merging. The range of $V_{m}$ observed at a smaller $Pr$ was larger
than that at a higher $Pr$, as could be noticed from the spread of the
pdfs $\square$ and $\blacktriangle$ in figure~\ref{rawpdf}. The higher
shear at lower $Pr$ could be the reason for such a behaviour of the
pdfs. The shear at smaller $Pr$ would have larger variations across
the planform since it has to be negligible near the side walls while
its value at the centre will increase obeying the standard relations
for the large scale flow
strength\cite*[][]{ahlers09:_heat_rayleig_benar_convec}; similar would be
the corresponding variation of $V_{m}$ since shear reduces $V_{m}$.

Comparing the pdfs at $Ra_w= 1.31\times 10^5$ and $2.65\times 10^8$,
which are approximately at the same $Pr$ ($\circ$ and $\square$ in
figure 16) it could be noticed that the mean merging
velocity 
increased with increase in $Ra_w$. Plumes merge faster with increase
in $Ra_w$ at the same $Pr$, presumably since more amount of heat has
to be transported into the bulk at a larger $Ra_w$. The range of
merging velocities also increase with increase in $Ra_w$, since due to
the same reason for the larger range of $V_{m}$ with lower $Pr$, the
range of variation of shear on the planform increases with $Ra_w$.
The pdf for $Ra_w=2.03\times 10^{11}$ and $Pr=602$ show the combined
result of the opposing effects of increase in $Ra_w$ and $Pr$. For
this data, the effect of an increase in $Pr$ seems to be dominant than
an increase in $Ra_w$ since the mean and the range of $V_{m}$ are
lower than those obtained at lower $Ra_w$.

The noticeable common trend from all the curves in figure \ref{rawpdf}
is that an increase in mean merging velocity decreases the probability
of finding this $\overline{V_{m}}$ in any planform; the curves are
shifted downward with increasing values of $\overline{V_{m}}$ for any
fluid. This trend is seen whether the increase in $\overline{V_{m}}$
occurs due to a decrease of $Pr$ or due to an increase of $Ra_w$. We
expect the reason for this behaviour to be as follows: The total
length of plumes in the planform $L_p$ increases as $Ra_w^{1/3}$ and
$Pr^{-0.1}$ as shown by \cite{puthenveettil11:_lengt}. Since
$\overline{V_{m}}$ increases with an increase in $Ra_w$ or a decrease
of $Pr$, an increase in $\overline{V_{m}}$ will also be accompanied by
a larger $L_p$ in the planform.  When $L_p$ increases, we expect the
total number of merging occurrences, as well as the number of mergings
with a specific $V_{m}$ also to increase. But since this increased
number of mergings occur over a larger range of $V_{m}$, as we saw
earlier, the probability of finding any $\overline{V_{m}}$ decreases
with an increase in $\overline{V_{m}}$.
\begin{figure}
  \centering
  \subfigure[]{\includegraphics[width=0.33\textwidth]{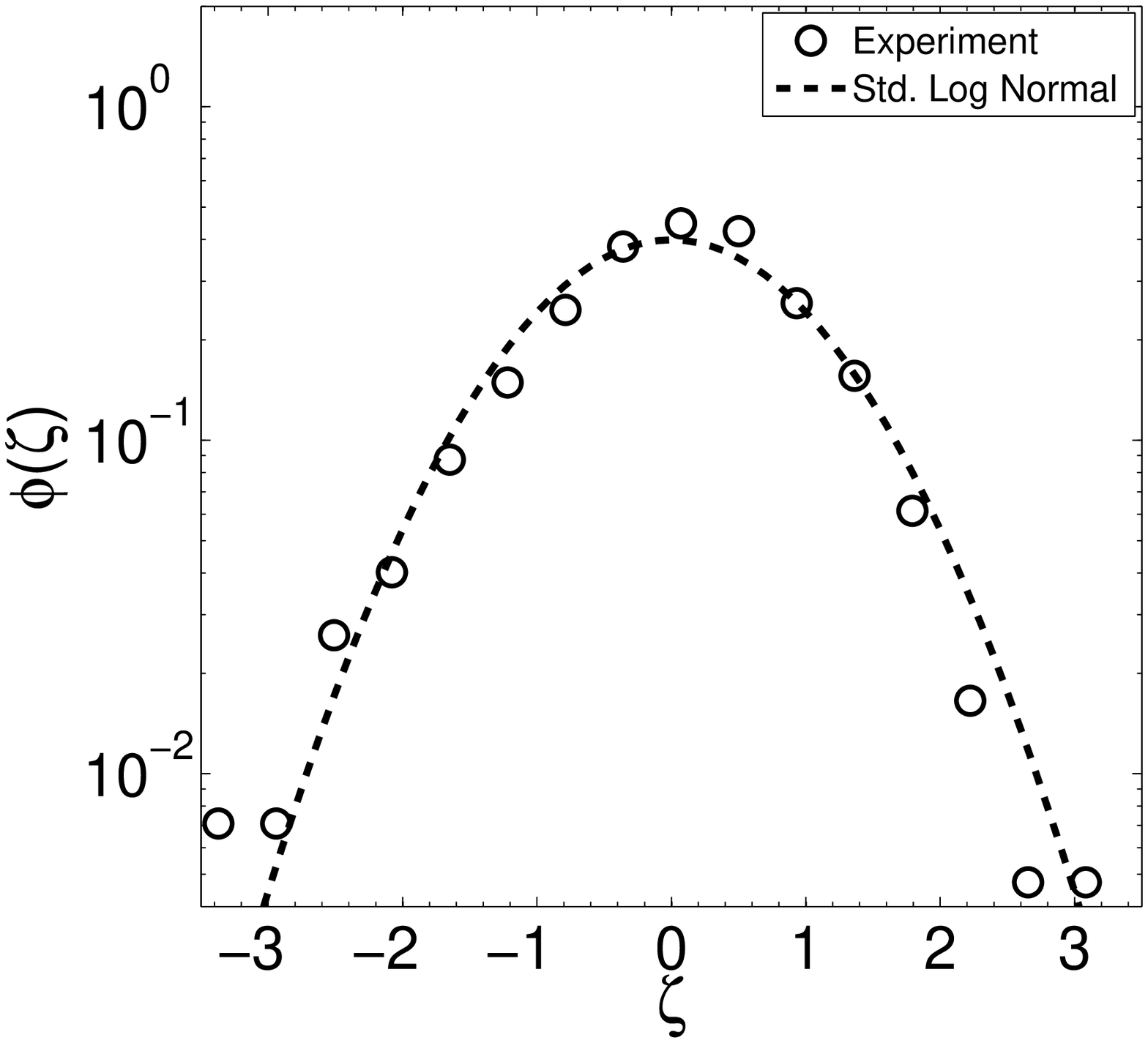}\label{commonpdf}}\hfill
 \subfigure[]{\includegraphics[width=0.33\textwidth]{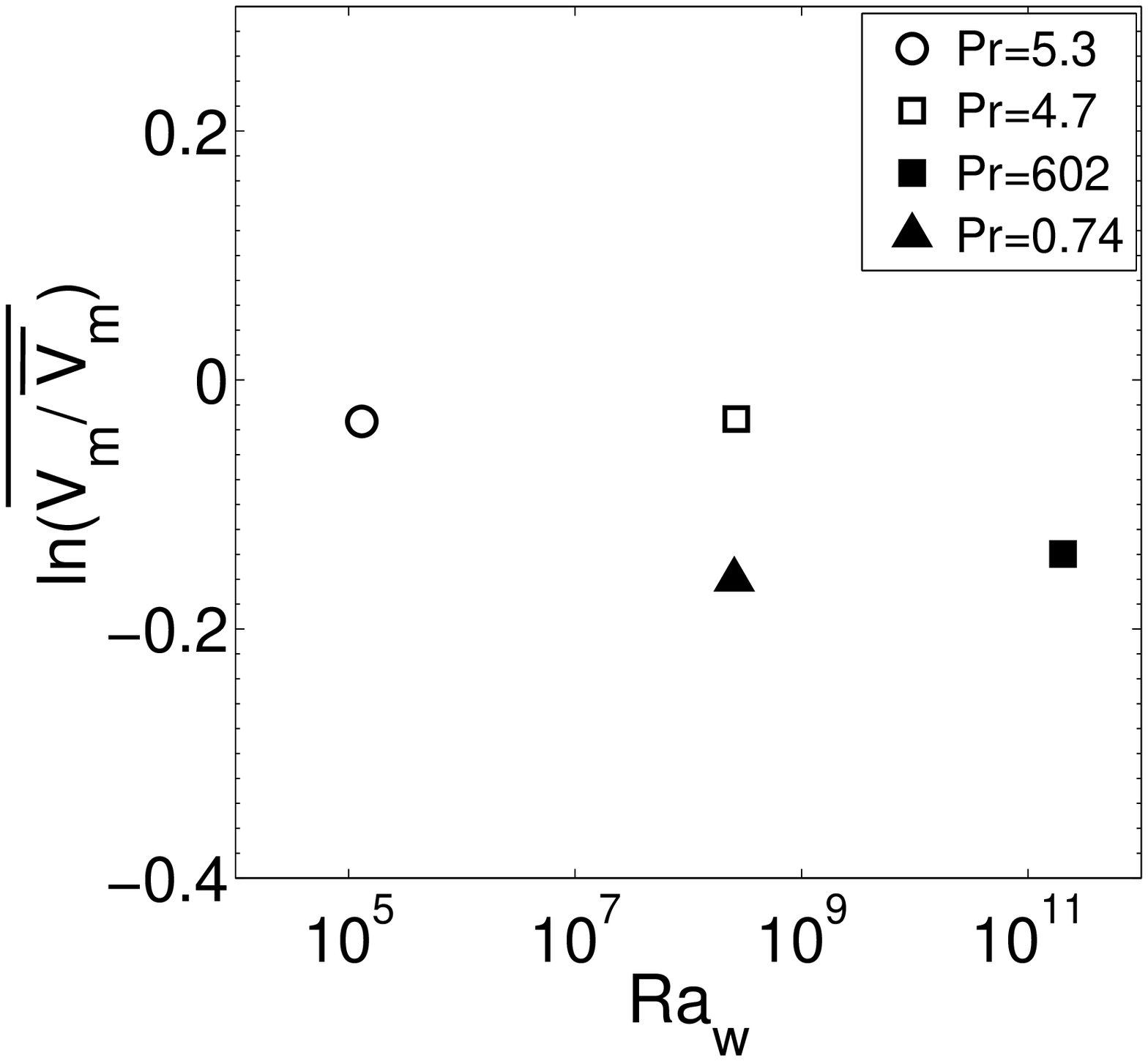}\label{lnmean}}\hfill
  \subfigure[]{\includegraphics[width=0.33\textwidth]{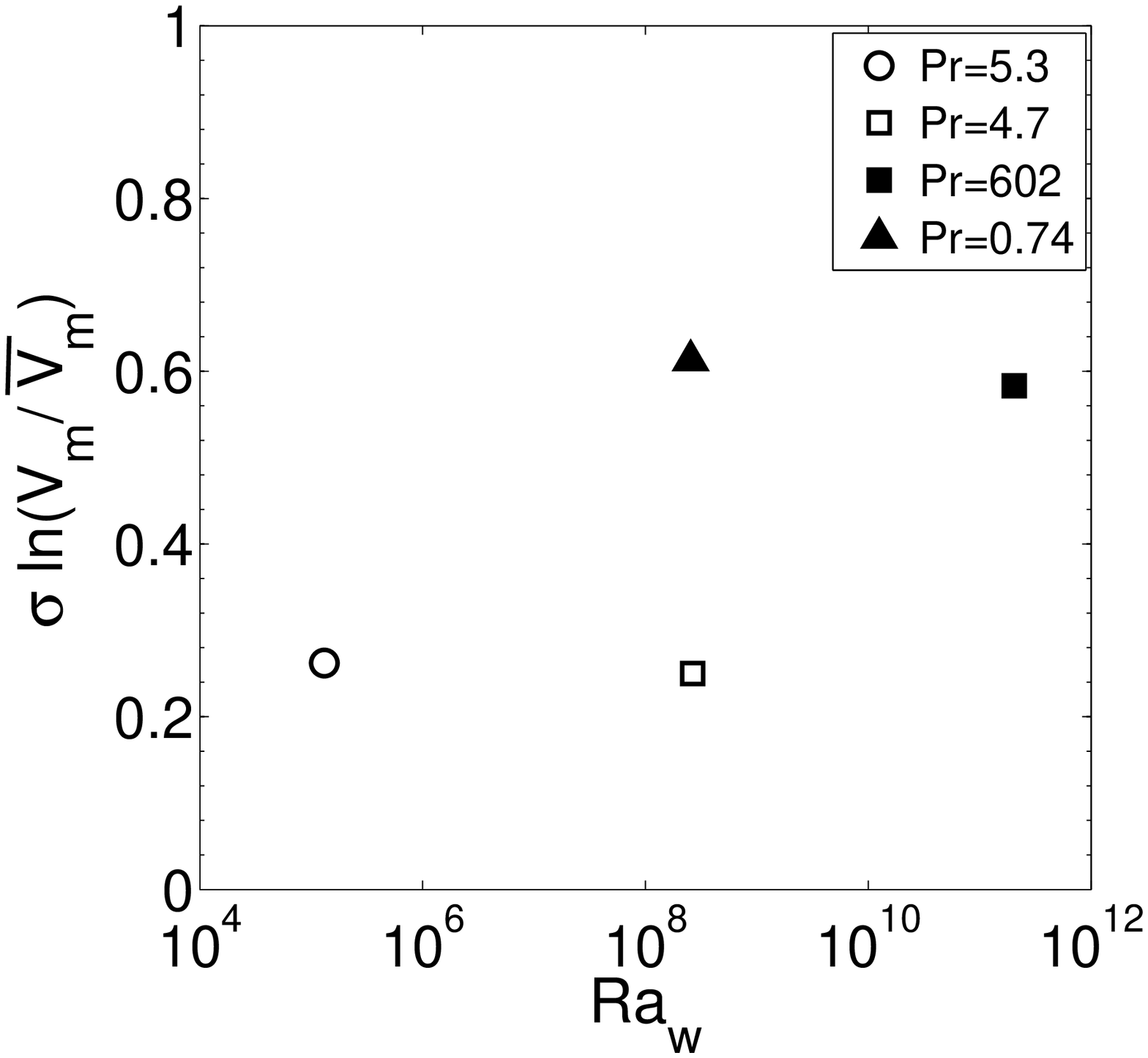}\label{lnstd}}
  \caption{(a), The common probability density function of the logarithm of
   the dimensionless plume merging velocities in their standardised form $\zeta=
(\ln(V_{m}/\overline{V_{m}})-\overline{\ln(V_{m}/\overline{V_{m}})})/\sigma\ln(V_{m}/\overline{V_{m}})$ at all $Ra_{w}$ and $Pr$; (b), Variation of the mean of the logarithm of
    the dimensionless plume merging velocities with $Ra_w$; (c),
    Variation of the standard deviation of the logarithm of the
    dimensionless plume merging velocities}
  \label{fig:pdf}
\end{figure}

The inset in figure \ref{rawpdf} shows the data in the main figure
plotted as the pdf of the logarithm of the dimensionless merging
velocities ($V_{m}/\overline{V_{m}}$) in its standardised form $\zeta=
(\ln(V_{m}/\overline{V_{m}})-\overline{\ln(V_{m}/\overline{V_{m}})})/\sigma\ln(V_{m}/\overline{V_{m}})$,
where, \textemdash$\,$ indicates the mean and $\sigma$ indicates the
standard deviation.  All the curves in figure~\ref{rawpdf} collapse on
to the standard log-normal curve, shown as the dashed line. Even
though the mean and the variance of $V_{m}$ increased with increasing
$Ra_{w}$ and decreasing $Pr$, the probability distribution function
has a common log-normal form for all the $Ra_w$ and $Pr$.  We hence
combine the merging velocity data, in the form of $\zeta$, for the
data shown in figure \ref{rawpdf} so as to obtain 1120 data points to
plot the general probability distribution function of plume merging
velocities in figure \ref{commonpdf}. The variation of the
standardising parameters $\overline{\ln(V_{m}/\overline{V_{m}})}$ and
$\sigma\ln(V_{m}/\overline{V_{m}})$ are shown in figures~\ref{lnmean}
and \ref{lnstd}. The unfilled symbols have a constant value of
$\overline{\ln(V_{m}/\overline{V_{m}})}\approx -0.025$ while the
filled symbols have a value about -0.14. Similarly,
$\sigma\ln(V_{m}/\overline{V_{m}})\approx 0.25$ for the unfilled
symbols while it is equal to 0.6 for the filled symbols. Studying
figure~\ref{rawpdf}, we observe that the pdf of the filled symbols,
which are either at a much larger $Ra_w$ or a much smaller $Pr$ than
the water experiments at $Pr\approx 5$, have a noticeable asymmetry
with a larger probability of finding the lower than mean values of
$V_{m}$ than the higher than mean values. This asymmetry lowers the
$\overline{\ln(V_{m}/\overline{V_{m}})}$ and increases the
$\sigma\ln(V_{m}/\overline{V_{m}})$ for the filled symbols, seen
respectively in figure~\ref{lnmean} and \ref{lnstd}, when compared to
the unfilled symbols. Shear increases with increase in $Ra_w$ and
decrease in $Pr$, and shear decreases the merging velocities. Hence at
larger $Ra_w$ and lower $Pr$ we expect a higher probability of
occurrence of $V_{m}$ smaller than the mean. Similarly, since increase
in shear also increases the range of $V_{m}$ we expect
$\sigma\ln(V_{m}/\overline{V_{m}})$ to be higher at higher $Ra_w$ and
lower $Pr$.
\subsubsection{Mean merging velocities}\label{sec:mean-merg-veloc}
\begin{figure}
  \centering
  \includegraphics[width=0.6\textwidth]{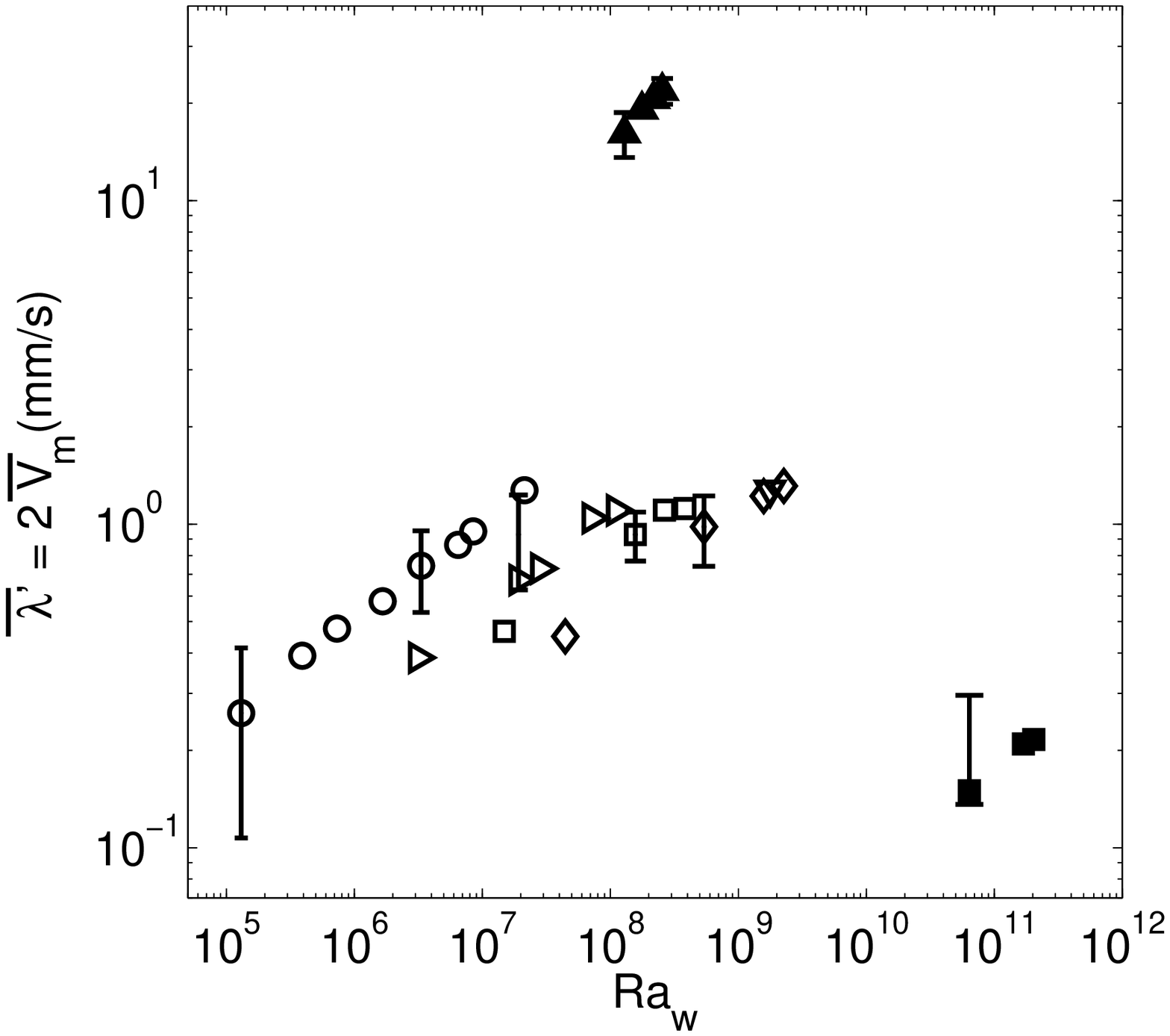}
  \caption{Variation of the mean plume merging velocity
    $\overline{V_{m}}$ with $Ra_{w}$ for the various $ Pr$. The open
    symbols show $\overline{V_{m}}$ for the thermal convection
    experiments in water at $Pr=3.6$ to $5.3$ for the following layer
    heights; $\circ,\, H= 50$mm; $\rhd,\, H= 100$mm; $\Box,\,
    H=150$mm; $\bigtriangledown,\,H=200$mm and $\diamond,\, H=210$mm.
    The variation of $\overline{V_{m}}$ for $Pr=0.74$ at $H=500$mm is
    shown by $\blacktriangle$ and for $Sc=602$ at $H=230$mm by
    $\blacksquare$.}
  \label{fig:meanmergvel}
\end{figure}

Figure(\ref{fig:meanmergvel}) shows the variation of
$\overline{V_{m}}$, the mean velocity of merging of plumes, with
$Ra_{w}$ for the three $Pr$; the corresponding values are listed in
table~\ref{tab:parameter}.  At each $Ra_{w}$, the mean was determined
from five values of plume merging velocities $V_{m}$ measured at
different locations, including shear and no-shear regions, at
different times.  The vertical bars give the range of $V_{m}$ obtained
from such measurements; an indication of the range of $V_{m}$ at any
$Ra_w$ due to varying shear at different locations and at different
times.  As seen earlier, the effect of shear is to reduce the plume
merging velocity \cite*[][]{puthenveettil05:_plume_rayleig}. The range
of velocities at $Ra_{w}=1.31 \times 10^{5}$ and $Pr=5.3$ shows that
$\overline{V_{m}}$ is closer to the maximum value implying that more
measurements were made from low shear regions. On the contrary, the
$\overline{V_{m}}$ measured at $Ra_{w}=6.39 \times 10^{10}$ and
$Pr=602$ is closer to the minimum value of $V_{m}$ indicating more
measurements from shear dominated regions. As seen in
\S~\ref{sec:shear-along-plumes}, with increase in $Ra_w$, in addition
to increase in shear, the regions affected by shear also increases;
more and more $V_{m}$ in the planform are affected by shear with
increase in $Ra_w$, which reduces its value.  The measurements of
$V_{m}$ hence seems to account for the effect of shear due to the
large scale flow.

At all $Pr$, for any constant $H$, $\overline{V_{m}}$ increased with
increase in $Ra_{w}$; plumes merged faster with increase in driving
potential $\Delta T_{w}$. Variation of $\overline{V_{m}}$ for thermal
convection experiments in water at approximately the same $Pr$, shown
by any of the symbols $\circ,\rhd, \Box$ and $\diamond$ in figure
\ref{fig:meanmergvel}, were obtained by changing the heat flux for any
constant $H$. It is clear that a change in layer height results in a
horizontal shift of the curve of $\overline{V_{m}}$ without changing
the values of $\overline{V_{m}}$; as we show later, $\overline{V_{m}}$
is independent of layer height if the effect of external shear is
small.  The shift in $\overline{V_{m}}$ vs $Ra_w$ in figure
\ref{fig:meanmergvel} comes only due to the plotting of
$\overline{V_{m}}$ as a function of $Ra_w$ by which the value of
$Ra_w$ at the same driving potential is shifted by a fixed value
between experiments with different layer heights.  Comparing the water
and air data at $Ra_w\sim 5\times 10^8$ it could be noticed that
the plumes merged much faster at $Pr=0.74$ than at $Pr=5.2$.
$\overline{V_{m}}$ is inversely proportional to $Pr$, as was also
observed in figure \ref{rawpdf}. By comparing
figures~\ref{fig:meanmergvel} and \ref{fig:meanshear}, we also notice
that the mean merging velocities are an order lower than the mean
longitudinal velocities. We now quantify these qualitative
observations about the mean merging velocities by building a scaling
analysis based on an assumed phenomenology of merging laminar sheet
plumes.
\section{Scaling of mean merging velocities}
\label{sec:scaling-mean-merging}
\begin{figure}
  \centering
  \subfigure{\includegraphics[width=0.465\textwidth]{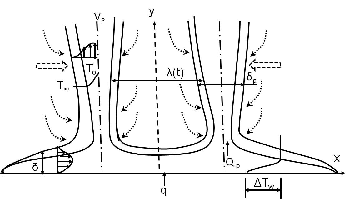}\label{fig:schemplummerg}}\hfill
  \subfigure{\includegraphics[width=0.525\textwidth]{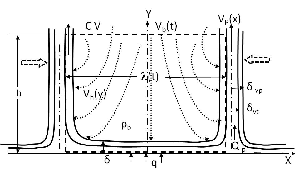}\label{fig:controlvolume}}
  \caption{(a), Schematic of two adjacent line plumes on the plate
    that are merging at any instant, separated by the plume spacing
    $\lambda(t)$.  The hollow arrows show the direction of merging and
    the curved arrows the entrainment into the plumes. (b), The
    idealisation used for the scaling analysis of the mean merging
    velocities $\overline{V_m}$. The deforming control volume (CV) of
    height of $h$ and width $\lambda(t)$ is shown as the dashed
    line. $V_{b}(t)$ is the uniform velocity of the bulk fluid into
    the CV at the top surface of the CV at any instant, $V_{h}(y)$ is
    the velocity of the fluid entrained into the plume at the edge of
    the plume, $V_p(x)$ is the velocity distribution inside the plume
    at a height of $h$. The thickness of the velocity boundary layer
    feeding either side of a plume is $\delta_{v}$ while the plume
    boundary layer thickness is $\delta_{vp}$. $q$ is the total heat
    flux supplied to the plate and $Q_{p}$ is amount of heat
    transported by the plume per unit length at any height.}
 \label{fig:schemplummergcv}
\end{figure}

Consider two parallel line plumes that are merging as shown in
figure~\ref{fig:schemplummerg}. These line plumes originate from the
instability of laminar natural convection boundary
layers~\cite*[][]{pera73} that occur on both sides of each plume at
its base,
\citep{puthenveettil05:_plume_rayleig,puthenveettil11:_lengt}. Since
these plumes are an outcome of the instability of the laminar natural
convection boundary layers, we expect these plumes to retain their
laminar nature till some height. The similarity solutions of such
laminar plumes rising from a line source of heat are given by
\cite{fuji63:_theor} and ~\cite*[][]{gebhart70:_stead}. The rising
plumes entrain fluid from the ambient through their sides so that the
mass flux at any cross section, the velocities inside the plumes and
the width of the plumes increase with height
~\cite*[][]{gebhart70:_stead}. The entrainment flow at the side of the
plume at any height is determined by the velocity inside the plumes,
which again is determined by the heat flux. For a given heat flux and
spacing between the plumes, the flow from the bulk region may not be
enough to meet this entrainment flow into the plume; the plumes could
then be expected to adjust their horizontal position so that mass and
momentum balance of the region between the plumes is maintained. We
formalise this phenomenology below by using the unsteady mass and
momentum balance of a deforming control volume ( CV ) in between the
plumes.
\subsection{Control volume mass and momentum balance}
\label{sec:mass-moment-balance}
As shown in figure~\ref{fig:controlvolume} we choose the width of the
CV to be equal to the spacing between the plumes $\lambda(t)$ which is
a function of time $t$. The height $h$ of the CV is chosen as the
height to which the diffusive region near the plate extends upwards.
$h=a Z_w$ is chosen as the position at which the mean profiles of
temperature, velocity and their fluctuations do not show substantial
variation with height ( see
\cite{theerthan00:_planf,theerthan98:_rayleig}).  Here, for the time
being, $a$ is assumed to be a constant and $Z_w$ is the length scale
defined by (\ref{eq:zw}). In actuality, $a$ is a function of $Pr$; we
will see later that this has to be so for our theory to match the
experiments in its $Pr$ dependence. $a \approx 40$ for $Pr\sim 1$ so
that for the usual range of $Ra_w$, $h< 1$ cm. Over this small height,
at a scaling level, one could neglect the variation of the plume
thickness with height in estimating the volume of fluid in between the
plumes. Further since $\delta_v\ll h$, we neglect the contribution of
the boundary layers on the plate in estimating the volume of the fluid
in between the plumes. Under these assumptions, the mass balance
becomes,
\begin{equation}
  \label{net mass flux}
2 \rho_{b}h V_m\approx\rho_{b}V_{b}\lambda-2\rho_{b}\int_{0}^{h}V_{h}dy,
\end{equation}
where, as shown in figure \ref{fig:controlvolume}, $\rho_b$ is the
density of the bulk fluid, $V_b(t)$ the downward velocity of the bulk
fluid at the top surface of the CV and $V_h(y)$ the horizontal
entrainment velocity at the edge of the plumes at any height
$y$. Equation (\ref{net mass flux}) shows that the rate of change of
mass of the bulk fluid in between the plumes due to the merging of the
plumes is equal to the difference between the influx from the bulk and
the out flux due to the entrainment by the plumes. Since $V_h(y)$
could be obtained as a function of $Ra_w$ and $Pr$ from the similarity
solutions of \cite{gebhart70:_stead}, (\ref{net mass flux}) could be
used to obtain the scaling of $\lambda^\prime$ if $V_b$ is known, or
could be eliminated by another equation. The momentum balance of the
fluid in between the plumes gives us such an equation.

By vertical momentum balance in the CV, we get,
\begin{equation}
  \label{momentum balance}
  \frac{\partial}{\partial t}\left(\rho_b\lambda\int_{\delta_v}^hVdy\right)\approx \rho_{b}gh\lambda-\Delta P\lambda+\rho_bV_b^2\lambda,
\end{equation}
under the same assumptions used in obtaining (\ref{net mass
  flux}). Here, $V(y)$ is the vertical velocity distribution in
between the plumes and $\Delta P$ the pressure difference between the
bottom and the top of the CV. The pressure drop across the boundary
layer on the plate ($\sim g\Delta\rho\delta_T$, see \S~A.2 in
\cite{puthenveettil11:_lengt}) is neglected in comparison to the
pressure drop in the rest of the height of the CV, $\rho_b
g(h-\delta_T)$. The above equation shows that the rate of change of
vertical momentum in between the plumes will be equal to the sum of
the net vertical force and the influx of vertical momentum due to the
inflow of bulk fluid at the top of the CV. We now estimate the unknown
$\Delta P$ in (\ref{momentum balance}) using unsteady Bernoulli
equation.

The flow in between the plumes could be assumed to be irrotational,
except in the boundary layer region, as was shown by
\cite{schneider81:_flow} for the case of axisymmetric plumes. We
neglect the pressure drop across the boundary layers on the plate and
the entrainment velocity at the edge of the boundary layers on the
plate in comparison to the velocity at the top of the CV, and assume
that $h-\delta_v\approx h$. Then, by applying unsteady Bernoulli
equation between the top of the CV and at the edge of the boundary
layer, we obtain,
\begin{equation}
  \label{pressure difference}
  \Delta P\approx\frac{1}{2}\rho_bV_b^2+\rho_bgh-\rho_b\frac{\partial}{\partial t}\int_{\delta_v}^{h}Vdy.
\end{equation}
With the above three equations (\ref{net mass flux}), (\ref{momentum
  balance}) and (\ref{pressure difference}) we now relate the unknown
merging velocity $V_m$ to the entrainment velocity $V_h$, which could
be estimated from the similarity solutions of \cite{gebhart70:_stead}.
\subsection{Relationships of merging velocity to velocities of
  entrainment and plume rise}
\label{sec:relat-merg-veloc}
Expanding the first term in (\ref{momentum balance}) and substituting
$\Delta P $ from (\ref{pressure difference}) in (\ref{momentum
  balance}), we obtain,
\begin{equation}
  \label{y mom bal equation}
  \frac{V_b\lambda}{h}\approx
  4V_m\left[\frac{\int_{\delta_v}^hVdy}{V_bh}\right]\sim 4V_m,
\end{equation}
since $\int_{\delta_v}^hVdy$ scales as $V_bh$. Now, the unknown $V_b$
in (\ref{net mass flux}) could be eliminated by substituting (\ref{y
  mom bal equation}) in (\ref{net mass flux}) to obtain,
\begin{equation}
  \label{plumemerg_vel}
  V_m\sim \frac{1}{h}\int_{0}^{h}V_h dy.
\end{equation}
In other words, the lateral merging velocity of the plumes $V_m$ is
equal to the average entrainment velocity at the sides of the plumes,
where the averaging is done over the height of the diffusive zone near
the plate.

We could also relate the merging velocity of plumes to their rise
velocity as follows. By mass balance of the plume,
\begin{equation}
  \label{eq:vhtovp}
  2\int_0^hV_h dy=\int_{-\delta_{vp}}^{\delta_{vp}} V_p dx,
\end{equation}
where $V_p(x)$ is the velocity distribution inside the plume across
the thickness of the plume of $2\delta_{vp}$ at a height of $h$.
Equations (\ref{plumemerg_vel}) and (\ref{eq:vhtovp}) would then imply
that
\begin{equation}
  \label{eq:lambdaprimevp}
  V_m\sim \overline{V_p}\, \frac{\delta_{vp}}{h},
\end{equation}
where $\overline{V_p}$ is the average plume rise velocity at a height
of $h$. Since $\delta_{vp}\ll h $, equation (\ref{eq:lambdaprimevp})
implies that the merging velocities of the plumes are an order lower
than the plume rise velocities. Since the velocity of large scale
circulation also scales as the velocity of rise of
plumes~\cite*[][]{puthenveettil05:_plume_rayleig}, the merging
velocities of the plumes would be an order lower than the large scale
circulation velocity as was observed figures~\ref{fig:meanmergvel} and
\ref{fig:meanshear}. We observed in \S~\ref{sec:vari-merg-veloc} that
the merging velocities are a constant during any merging instance at a
fixed $Ra_w$. It is easy to understand the cause for this behaviour
from (\ref{plumemerg_vel}) and (\ref{eq:lambdaprimevp}). At any
$Ra_w$, the strength of two adjacent plumes could be expected to be
constant over their merging period. This would imply that their
average entrainment velocity also remains constant during the merging
cycle;~(\ref{plumemerg_vel}) would then imply that the merging
velocities are also constant.
\subsection{The entrainment flow into the plumes}
\label{sec:plume-ent-veloc}
We have seen in (\ref{plumemerg_vel}) that the velocity of merging of
plumes is related to $\int_{0}^{h}V_h dy$, the total entrainment flow
into the plume through one of its sides, over the height of the
diffusive region near the plate. We now need to estimate this
entrainment flow as a function of $Ra_w$ and $Pr$ to find the scaling
of merging velocities. As given by \cite{gebhart70:_stead}, the
velocity of entrainment at the edge of a two dimensional plume above a
heated line source at any height $y$ is
\begin{equation}
  \label{horplumvel}
  V_h(y)= \frac{3\times 4^{3/4}}{5} \frac{\nu}{y} Gr_y^{1/4} f(\eta_{\delta_{vp}}),
\end{equation}
where, $\eta_{\delta_{vp}}$ is the value of the similarity variable
$\eta=(Gr_y/4)^{1/4}x/y$ of \cite{gebhart70:_stead} at the velocity
boundary layer thickness $\delta_{vp}$. $f(\eta)$ is the dimensionless
stream function defined by $\psi=4\nu(Gr_y/4)^{1/4}f(\eta)$, with the
local Grashoff number,
\begin{equation}
  \label{eq:localgr}
  Gr_y=g\beta(T_0-T_\infty)y^3/\nu^2.
\end{equation}
$f(\eta_{\delta_v})=a_1Pr^{-b_1}$ where $b_1=0.4$ for $Pr<1$ and $0.2$
for $Pr>1$, while $a_1=0.8$ for all $Pr$ as obtained from
\cite{gebhart70:_stead}. Here, $T_0-T_\infty$ is the temperature
difference between the center line of the plume and the ambient, which
decreases with $y$ as follows
\begin{equation}
  \label{eq:t0-tinfy}
  T_0-T_\infty=\left(\frac{Q_p}{\rho C_p I}\right)^{4/5}\left[\frac{1}{4^3g\beta\nu^2y^3}\right]^{1/5},
\end{equation}
where $I= 1/\sqrt{Pr}$ from \cite{gebhart70:_stead}, $C_p$ is the
specific heat at constant pressure and $Q_p$ is the heat flux into
each line plume. Equation~(\ref{horplumvel}) is for a single plume
above a heated line source, in the case of convection over a hot
surface, we need to relate the heat flux $q$ from the plate to the
strength of the line source $Q_p$ of individual plumes in
(\ref{eq:t0-tinfy}).
\subsubsection{Relation between $Q_p$ and $q$}
\label{sec:relation-between-q_p}
The heat from the plate is transferred to the boundary layers on both
sides of each of the plumes, which then input this heat to the base of
each of the plume.  The boundary layers, which become unstable at a
mean distance of $\overline{\lambda}$, cover the plate. Hence, we
assume that the total heat carried away by all the plumes from any
area $A$ on the plate at any instant is equal to the total heat
supplied by the plate in that area. Numerical studies by
\cite{shishkina08:_analy_rayleig} have shown that this assumption is
valid.

$Q_p$ in (\ref{eq:t0-tinfy}) is also the heat transported by the plume
per unit length at any height, since the total heat transported by the
plume is constant at any height. Now, if $L_p$ is the total length of
all the plumes in an area $A$ of the plate, then equating the heat
carried away by the plumes in an area $A$ to the heat supplied by the
plate in the same area $A$ implies,
\begin{equation}
Q_{p}L_{p}=qA.
\label{heatbal}
\end{equation}
Since
\begin{equation}
L_{p}=A/\overline{\lambda},
\label{totplulen}
\end{equation}
as shown by \cite{puthenveettil11:_lengt},
\begin{equation}
Q_{p}=q\overline{\lambda}.
\label{pluheat}
\end{equation}
\subsubsection{Relation between $Q_p$, $\overline{\lambda}$ and fluid properties }
To eliminate $Q_p$ from (\ref{eq:t0-tinfy}) , $q$ in (\ref{pluheat})
could also be written in terms of $\overline{\lambda}$ to obtain a
relation between $Q_p$, $\overline{\lambda}$ and fluid
properties. Using the expression for Townsend's near-wall length
scale,
\begin{equation}
Z_{o}=\left(\frac{\alpha^3 \rho C_p }{g \beta q}\right)^{1/4}
\label{fluxlenscale}
\end{equation}
and the relation,
\begin{equation}
\frac{\overline{\lambda}}{Z_{o}}= C_{2}Pr^{n_{2}},
\label{meanpluspaceflux}
\end{equation}
\cite*[][]{puthenveettil11:_lengt}, where, $C_{2}=31$ and
$n_{2}=0.345$, along with (\ref{pluheat}), we get
\begin{equation}
  \label{eq:Qp}
  Q_p=(C_2Pr^{n_2})^4\frac{\rho C_p}{g \beta} \left(\frac{\alpha}{\overline{\lambda}}\right)^3.
\end{equation}
We now substitute~(\ref{eq:Qp}) in (\ref{eq:t0-tinfy}) and use the
resulting expression for $T_0-T\infty$ in (\ref{eq:localgr}) to obtain
the expression for local $Gr_y$. Using this expression for $Gr_y$ in
(\ref{horplumvel}), we integrate the resulting expression for $V_h$
with respect to $y$ from $0$ to $h=aZ_w$. Replacing
$\overline{\lambda}$ in the resulting expression
with~(\ref{meanpluspace}) we obtain,
\begin{equation}
  \label{eq:lambdaprime}
  \int_{0}^{h}V_h\, dy\sim A_1Pr^{B} \nu
\end{equation}
where $A_1=\left(2^{6}C_2^4a^3/C_1^3\right)^{1/5}a_1 = 26$ and
$B=(4n_2-3n_1)/5-b_1-1/2= -0.7$ for $Pr<1$ and $B=-0.5$ for $Pr>1$.
Surprisingly, the total entrainment flow per unit length of line
plumes through the sides of the plumes over the height of the
diffusive region near the plate, given by (\ref{eq:lambdaprime}), is
independent of $Ra_w$ and is only a function of viscosity and $Pr$,
when the line plumes are separated by a mean spacing of
$\overline{\lambda}$.
\subsection{Scaling of $\lambda^\prime$}
\label{sec:scaling-lambdaprime}
Substituting (\ref{eq:lambdaprime}) in (\ref{plumemerg_vel}), we
obtain,
\begin{equation}
  \label{eq:vm}
  \overline{V_m}=APr^B \frac{\nu}{Z_w},
\end{equation}
where $\overline{V_m}$ is the mean merging velocity and
$A=A_1/a=0.65$.  We use $\overline{V_m}$ instead of $V_m$ in
(\ref{eq:vm}) since (\ref{pluheat}) is valid for a uniform array of
plumes separated by $\overline{\lambda}$. Equation~(\ref{eq:vm}) shows
that the mean lateral merging velocity of line plumes scales as the
velocity scale near the plate $\nu/Z_w$. Note that (\ref{eq:vm}) shows
that the merging velocity is only a function of the variables near the
plate.  This is however so because we have neglected the effect of
shear in the scaling analysis. As we have seen earlier in
\S~\ref{sec:vari-merg-veloc} and \ref{sec:lateral-motion-line}, shear
due to the large scale flow reduces the merging velocity of plumes; if
the effect of shear is not substantial, the above scaling is expected
to hold. The $Pr$ dependence of $\overline{V_m}$ shown in
figure~\ref{fig:meanmergvel}, where higher $Pr$ fluids had lower
merging velocities seems to be captured by the $Pr$ dependence in
(\ref{eq:vm}). The horizontal shift in $V_m$, observed in
figure~\ref{fig:meanmergvel}, between experiments with different
heights of fluid layer at the same $Pr$ could also be understood from
(\ref{eq:vm}). Since $V_m$ is not a function of $H$, at
least to first order, change in $H$ in the $Ra_w$ in the abscissa of
the plot will shift the data without changing its scaling. We now
quantitatively compare the scaling law of (\ref{eq:vm}) with the
experimental data of figure~\ref{fig:meanmergvel} in three different
dimensionless forms.
\subsubsection{Reynolds number relation}
\label{sec:reyn-numb-relat}
\begin{figure}
  \centering
  \includegraphics[width=0.7\textwidth]{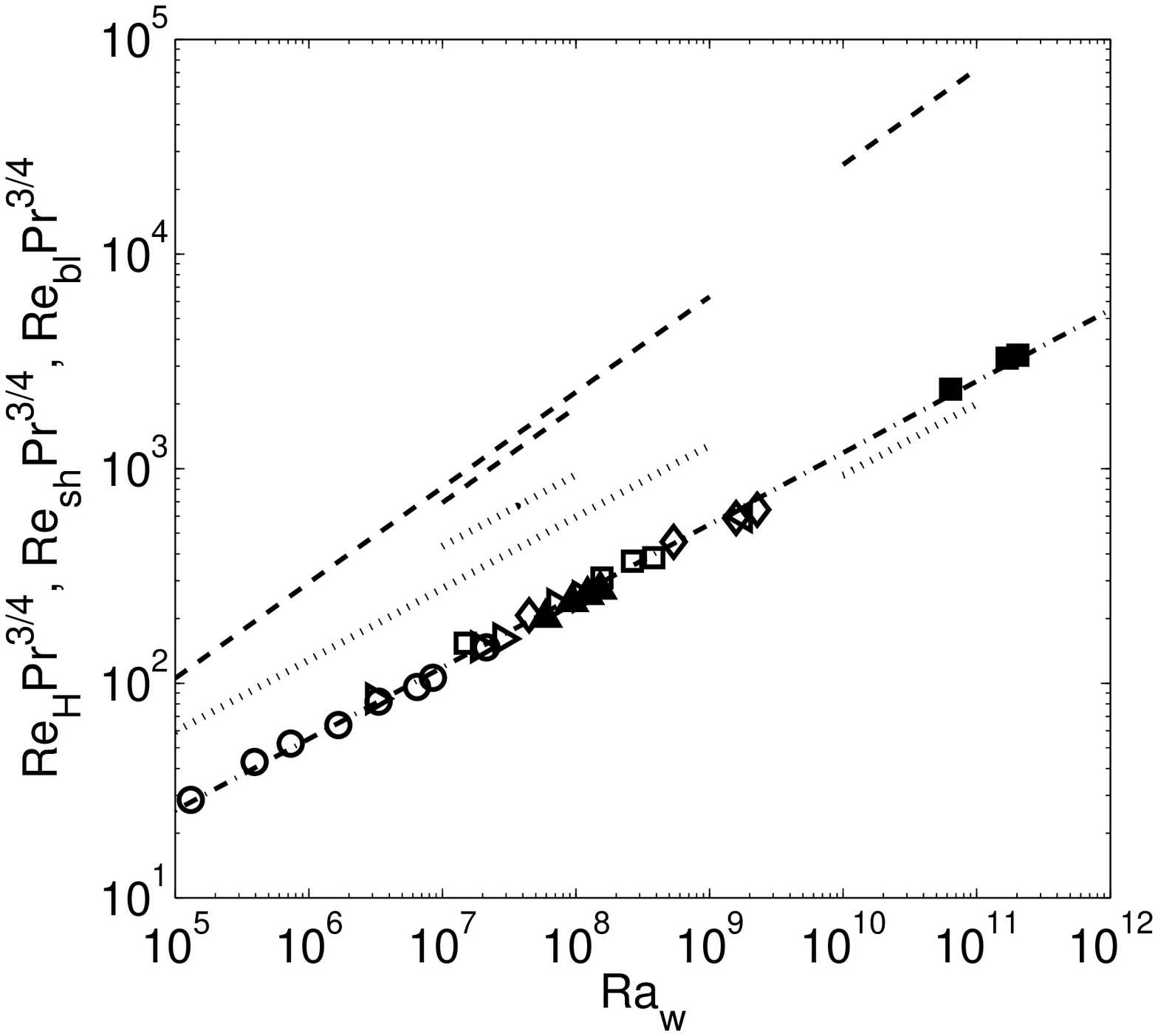}
  \caption{Variation of $Re_H= \overline{V_m}H/\nu$, the Reynolds
    number based on the mean merging velocity $\overline{V_m}$ and the
    layer height $H$, with $Ra_{w}$.  The open symbols indicate
    thermal convection experiments in water at $Pr=3.6$ to $5.3$ for
    the following layer heights; $\circ,\, H= 50$mm; $\rhd,\, H=
    100$mm; $\Box,\, H=150$mm; $\bigtriangledown,\,H=200$mm and
    $\diamond,\, H=210$mm.  $\blacktriangle$ represent experiments at
    $Pr=0.74$ and $H=500$mm while $\blacksquare$ represent experiments
    at $Sc=602$ and $H=230$mm. $-\cdot-\cdot\,
    Re_H=0.55Ra_w^{1/3}Pr^{-3/4}$~(\ref{rehexpt}); $ ---,
    Re_H=0.55Ra_w^{4/9}Pr^{-2/3}$~(\ref{eq:resh}) and
    $\cdot\cdot\cdot\cdot, Re_{bl}=1.9
    Ra_w^{1/3}Pr^{-0.98}$~(\ref{eq:rebl}) plotted for each of the $Pr$
    separately. }
  \label{fig:dimensionlessmergvel}
\end{figure}

The scaling law of (\ref{eq:vm}) could be rewritten as a Reynolds
number in terms of the mean merging velocity and the layer height,
$Re_H=\overline{V_m}H/\nu$ as follows,
\begin{equation}
 Re_{H}=A Ra_w^{1/3} Pr^B,
\label{reh}
\end{equation}
where $A=0.65$ and $B=-0.7$ for $Pr<1$ and $B=-0.5$ for $Pr>1$. As
shown in figure~\ref{fig:dimensionlessmergvel}, the measured mean
merging velocities, in terms of $Re_H$, scale as,
\begin{equation}
\label{rehexpt}
Re_{H} = 0.55 Ra_w^{1/3}  Pr^{-3/4},  
\end{equation}
at all $Pr$.  The $Ra_w$ dependence in the theoretical scaling of
$Re_H$ (\ref{reh}) matches exactly with the experiments for all $Pr$;
the difference in the prefactor between theory and experiments is also
small (0.65 as against 0.55). Same is the case with the exponent of
$Pr$ (-0.75 as against -0.7) for experiments with $Pr<1$. These
agreements are remarkable considering the number of approximations
that were made in the theory, the presence of spatially varying shear
in experiments which is not considered in the theory and other errors
in the experimental measurements.  Equation (\ref{reh}) and
(\ref{rehexpt}) imply that the mean plume merging velocity
$\overline{V_m}$, normalised by the viscous velocity scale $\nu/H$,
scales as $Ra_{w}^{1/3}$ for a given fluid. Such a scaling is similar
to the scaling of the dimensionless plume length $L_{P}/(A/H)$
and $Nu$ in turbulent convection. The dependence of $Re_{bl}$, the
Reynolds number in terms of the horizontal velocities inside the local
boundary layers at the base of the plume, given by (\ref{eq:rebl}), is
also shown in figure~\ref{fig:dimensionlessmergvel}; $Re_{bl}\gg
Re_H$. Since the velocities inside the boundary layers feeding the
plumes are larger than the merging velocities, the plumes are
sustained during their merging period.

Figure \ref{fig:dimensionlessmergvel} also shows the magnitude and
scaling of $Re_{sh}$ (\ref{eq:resh}) in comparison with that of
$Re_H$. As shown in \S~\ref{sec:shear-along-plumes}, $Re_{sh}$ has the
magnitude, and also scales as the large scale velocity $V_{LS}$.
Figure~\ref{fig:dimensionlessmergvel} shows that at all $Pr$, the magnitude of the
$\overline{V_{sh}}$ is an order of magnitude higher than that of the
mean merging velocities. The magnitudes of the mean velocities
obtained by \cite{hogg13:_reynol_prand} by spatial correlation of
images of plumes were substantially lower than that of the large scale
velocity. 
Since mean merging velocities are much smaller than the shear
velocities, the mean velocities detected by
\cite{hogg13:_reynol_prand} could be the mean merging velocities,
which could possibly be interpreted as following the scaling of large
scale flow over a smaller range of $10^5<Ra_w<10^7$ of
\cite{hogg13:_reynol_prand}.

For $Pr>1$, eventhough the theoretical scaling has the same prefactor
and dependance on $Ra_{w}$ as in the $Pr<1$ case, it has an exponent
of $Pr$ equal to $-0.5$ as against the experimental exponent of
$-0.75$. We now explore the possible cause of this discrepancy between
the theory and experiments for $Pr>1$.

The theoretical scaling law (\ref{reh}) was obtained by assuming
$h=aZ_w=aH/Ra_w^{1/3}$, where $a\approx 40$. However, this relation
for the height of the diffusive region near the plate is valid mostly
for $Pr\sim 1$ since $Pr$ dependence is not included in the length
scale $Z_w$ (see \cite{puthenveettil11:_lengt} where the boundary
layer thickness is a function of $Z_w$ and $Pr$). It is hence not
surprising that the $Pr$ dependence of the theory is correct for $Pr <
1$, since these experiments are at $Pr\approx 0.7$, close to
one. Since our $Pr>1$ experiments are at $Pr\approx 600$, much larger
than one, the $Pr$ dependence in $h$ should be accounted in the theory
if a better agreement with experiments is to be achieved at these
$Pr$. In addition, in obtaining the scaling shown by (\ref{y mom bal
  equation}), we had assumed that $\int_{\delta_v}^hVdy$ scales as
$V_bh$. Such a scaling is mostly correct only at $Pr\le 1$ since
viscosity could create vertical variations of vertical velocity in
between the plumes at larger $Pr$. Same is the case with the constant
$V_B$ across the top of the CV that was assumed in the analysis; at
large $Pr$, effect of viscosity could create horizontal variations of
the bulk flow into the region between the plumes. The above effects
are difficult to quantify since no solution for the far field of the
entrainment flow into a line plume is available, eventhough the
axisymmetric case has been solved by \cite{schneider81:_flow}. The
$Pr$ dependence of the height of the diffusive region near the plate
is hence not known as of now so as to get a better match of the theory
with the experiments.

It is easy to see that the availability of the above missing
information could result in an exact match of the theory with the
experiments. We notice that if the height of the diffusive region near
the plate is a function of $Pr$ so that,
\begin{eqnarray}
  a&=&60Pr^{1/8}\, \mathrm{for}\, Pr<1\,\mathrm{and}\label{eq:aprlt1}\\
  &=&60Pr^{5/8}\,\mathrm{for}\, Pr>1,\label{eq:aprgt1}
\end{eqnarray}
the theoretical scaling will match the experiments for all
$Pr$. Qualitatively, (\ref{eq:aprlt1}) and (\ref{eq:aprgt1}) seems to
be correct since, as judged from the distributions of mean quantities,
the height of the diffusive region near the plate increases with
increase in $Pr$ at the same $Ra_w$~\cite*[][]{theerthan98:_rayleig}.

Due to the above mentioned unavailability of information to make the
scaling analysis more rigorous, we assume that the experimental
scaling given by (\ref{rehexpt}) to be the more reliable one from
among (\ref{reh}) and (\ref{rehexpt}).  Alternatively one could assume
that the height of the diffusive region near the plate follows the
relation $h=aZ_w$, where $a$ is given by (\ref{eq:aprlt1}) and
(\ref{eq:aprgt1}), so that the scaling (\ref{rehexpt}) obtained from
the experiments match the theory. Further research needs to be
conducted to verify the assumed $Pr$ depends of the height of the
diffusive region near the plate. We now look at the implications of
(\ref{rehexpt}), interms of the invariants that come out of such a
scaling, as well relate the dimensionless flux $Nu$ to the
dimensionless merging velocity.
\subsubsection{Invariants}\label{sec:invariants}
\begin{figure}
  \centering
 \includegraphics[width=0.7\textwidth]{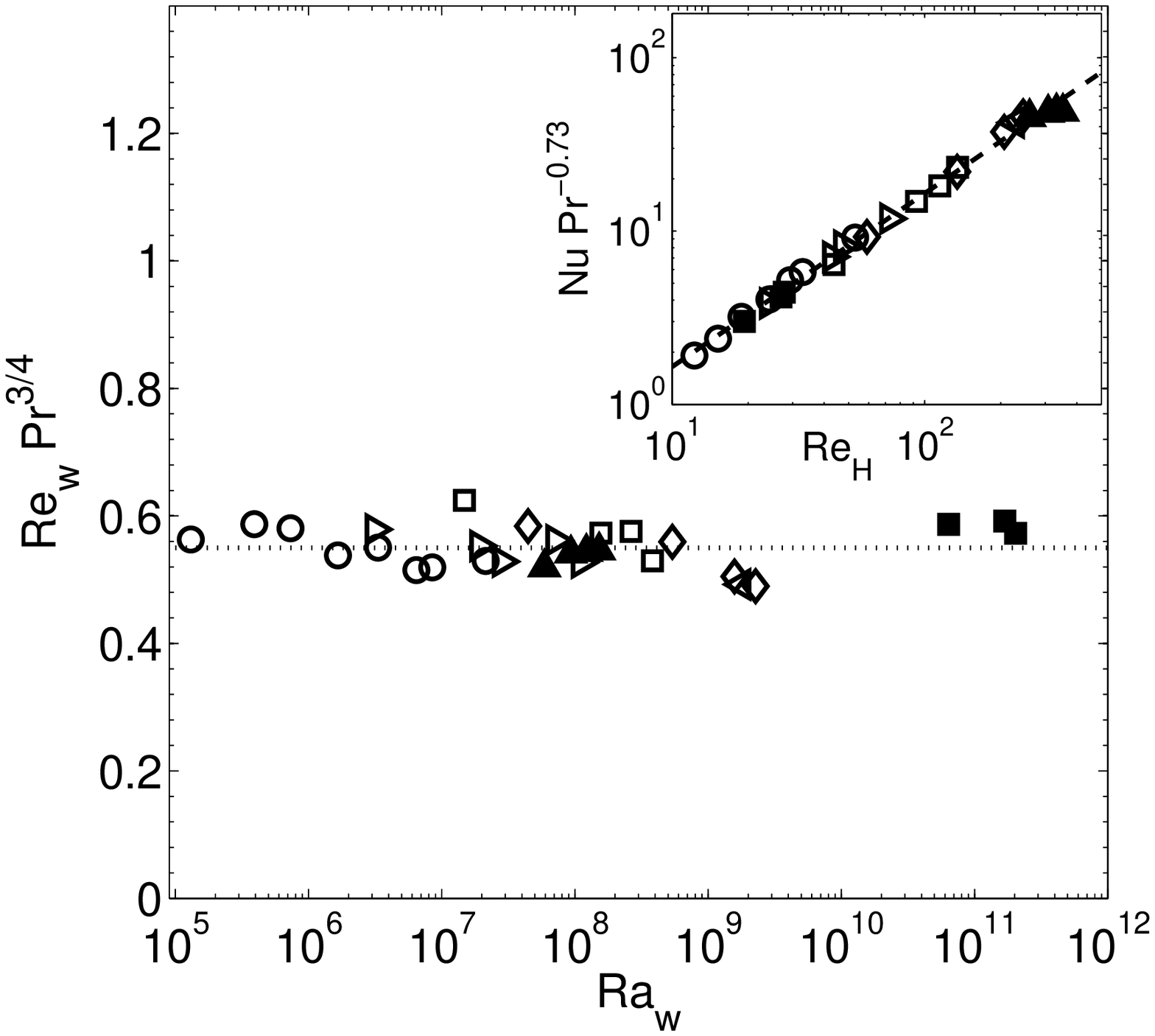}
 \caption{Variation of $Re_{w}=\overline{V_m}Z_{w}/\nu$, the Reynolds
   number based on variables near the plate, with $Ra_{w}$. The open
   symbols show experiments in water for $Pr=3.6$ to $5.3$ for the
   following layer heights; $\circ,\, H= 50$mm; $\rhd,\, H= 100$mm;
   $\Box,\, H=150$mm; $\bigtriangledown,\,H=200$mm and $\diamond,\,
   H=210$mm.  $\blacktriangle$ indicate experiments at $Pr=0.74$ and
   $H=500$mm while $\blacksquare$ show experiments at $Sc=602$ and
   $H=230$mm. $\cdot\cdot\cdot\cdot, Re_wPr^{3/4}=0.55$~(\ref{rew}).The inset shows
   the variation of Nusselt number $Nu$ with $Re_{H}$, the Reynolds
   number based on $V_{m}$ and layer height $H$; $---,
   Nu=0.165Re_{h}Pr^{0.73}$~(\ref{eq:rehnu}). }
  \label{fig:nearwallRe}
\end{figure}
Equation (\ref{eq:vm}) could be rewritten as an expression for a
Reynolds number based on the variables near the plate
$Re_w=\overline{V_m} Z_w/\nu$ $=0.55Pr^{-3/4}$. In other words,
\begin{equation}
\label{rew}
Re_wPr^{3/4}=0.55
\end{equation}
is a dimensionless invariant for all the fluids in turbulent
convection. Figure \ref{fig:nearwallRe} shows that (\ref{rew}) holds
reasonably well for $0.7\le Pr\le 600$ and for $1.31\times 10^5\le
Ra_w \le 2\times 10^{11}$ in our experiments. Equation (\ref{rew})
means that, the mean merging velocity of the plumes normalised by the
velocity scale near the plate $\nu/Z_w$ is only a function of Pr, and
hence remains an invariant for any given fluid in turbulent
convection. If we define a Froude number
$Fr_w=\overline{V_m}/\sqrt{gZ_w}$ in terms of the mean merging
velocity and the length scale $Z_w$ near the plate, then,
\begin{equation}
  \label{eq:frre}
  Re_w^2Pr=\frac{Fr_w^2}{\beta \Delta T_w}.
\end{equation}
Using (\ref{eq:frre}) and (\ref{rew}) we obtain that,
\begin{equation}
  \label{eq:frinv}
  \frac{Fr_w^2}{\beta\Delta T_w\sqrt{Pr}}=0.3
\end{equation}
is another dimensionless invariant for the dynamics of plumes near the
plate in turbulent convection.
\subsubsection{Nusselt number relation}\label{sec:nuss-numb-relat}
$\overline{\lambda}$ given by (\ref{meanpluspace}) is also equal to
that given by (\ref{meanpluspaceflux}), thereby implying that,
\begin{equation}
  \label{eq:zwzo}
  Z_w=\frac{C_2}{C_1}Pr^{n_2-n_1} Z_o,
\end{equation}
where $Z_o$ is given by (\ref{fluxlenscale}). We replace the heat flux
that appears in $Z_o$ in (\ref{eq:zwzo}) in terms of the Nusselt
number $Nu=q/( k\Delta T/H)$.  Substituting the resultant expression
for $Z_w$ in (\ref{rew}), we obtain
\begin{equation}
  \label{eq:rehnura}
  Re_H=\frac{C_1}{2C_2}Pr^{1+n_1-n_2}(NuRa)^{1/4}.
\end{equation}
Replacing $Ra$ in (\ref{eq:rehnura}) by the expression for $Ra$
obtained from (\ref{rehexpt}), we obtain,
\begin{equation}
  \label{eq:rehnu}
  Nu=0.165Re_HPr^{0.73}.
\end{equation}
The inset of figure~\ref{fig:nearwallRe} shows that the experimental
$Nu$ and $Re_H$ obey the above expression satisfactorily.  $Re_H$ is
directly proportional to the Nusselt number, which should also be
obvious from (\ref{rehexpt}) since $Nu\sim Ra^{1/3}$ in turbulent
convection. However, it should be noted that (\ref{eq:rehnu}) has been
obtained here by not using any assumed flux scaling relation in
turbulent convection. We have only used the relation
(\ref{meanpluspaceflux}) for the mean plume spacing, which could be
obtained by the phenomenology of laminar natural convection boundary
layers becoming unstable at a critical thickness, as was shown by
\cite{puthenveettil11:_lengt}.
\section{Conclusions}\label{sec:conclusion}
In the present study of the dynamics of line plumes on the bottom
plate in turbulent convection, we identified longitudinal motion,
lateral merging and initiation as points as the predominant types of
motion of these line plumes. The primary conclusions about these
three types of motion of the plumes are as follows.
\begin{enumerate}
\item The magnitude and scaling of the mean longitudinal motion along
  the plumes ($\overline{V_{sh}}$) at any instant is same as that of
  the large scale flow.
\item The mean lateral merging velocity ($\overline{V_m}$) at any
  instant scales as the velocity scale $\nu/Z_w$ near the plate and
  are an order smaller than the large scale velocity.
\item The mean time for initiation of plumes $\overline{t^*} \sim
  Z_w^2/\alpha$, the diffusive time scale near the plate.
\end{enumerate}

These conclusions were obtained from experiments in thermally driven
turbulent convection in air ($Pr=0.7$), water ($Pr= 3.6$ to $5.3$) and
in concentration driven convection in water ($Sc=600$) over a six
decade range of Rayleigh numbers ($10^5<Ra_w<10^{11}$). The velocities
of the plumes were measured from visualisations in a horizontal plane
close to the bottom plate, by laser scattering of smoke particles,
electrochemical method and PLIF for the three $Pr$ experiments
respectively.

Among the three major types of motion of plumes, merging seems to
occur in a large fraction of the area of the plate for the range of
$Ra_w$ and $Pr$ in our experiments. The fraction of the length of the
plumes that undergo merging ($L_{pm}/L_p$) reduced from about 90\% at
$Ra_w\approx 10^5$ to about 55\% at $Ra_w\approx 10^9$; these
fractions becoming larger with increase in $Pr$~(\ref{eq:lpmlp}). We
found that $L_{pm}/L_p$ scales as $Ra_w^{-0.04}
Pr^{0.1}$~(\ref{eq:lpmlp}). Interestingly, the fraction of the plume
length affected by shear had an exact inverse dependence on $Ra_w$ and
$Pr$~(\ref{eq:lpslp}), so that the product of the lengths of plumes
that merge and are affected by shear is a constant fraction of the
square of the total plume length~(\ref{eq:lpslpm}).

The plumes merge all over the planform with a range of velocities so
that the merging velocities ($V_m$) show a distribution at any $Ra_w$
and $Pr$. Such a distribution is partly caused by the spatially and
temporally varying shear field near the plate due to the large scale
flow, since shear reduces the merging velocities. We found that the
distribution of merging velocities have a common log-normal
distribution at all $Ra_w$ and $Pr$; however the mean and standard
deviations of the distributions increased with increase in $Ra_w$ and
decrease in $Pr$~(\S~\ref{sec:stat-merg-veloc}). Plumes merged faster,
with their merging velocities spread over increasing ranges, with
increase in $Ra_w$ and decrease in $Pr$. Significantly, each merging
instance of adjacent plumes occurs with a constant velocity. As we
show from a control volume balance, such constant $V_m$ occurs since
$V_m$ is proportional to the average entrainment velocity
($\overline{V_h}$) at the sides of the plumes ~(\ref{plumemerg_vel}).
During a merging cycle at a given $Ra_w$, the strength of two adjacent
plumes remains constant, and hence $\overline{V_h}$ which is
proportional to the plume rise velocities, too is a constant.

Using similarity solutions of \cite{gebhart70:_stead} for estimating
$\overline{V_h}$ and relations for the mean plume spacing, to relate
heat flux into the plumes to the heat flux into the plate, we obtained
scaling laws for $\overline{V_m}$, the mean merging
velocities. $\overline{V_m}\sim \nu/Z_w$~(\ref{eq:vm}), the
appropriate velocity scale near the plate, where $Z_w$~(\ref{eq:zw})
is the relevant length scale near the plate in turbulent
convection. The Reynolds number in terms of $\overline{V_m}$ and layer
height $H$ scales as $Ra_w^{1/3}$ (\ref{rehexpt}) or as the
dimensionless flux $Nu$~(\ref{eq:rehnu}). The above relations also
implied that $\overline{V_m}$ made dimensionless by the appropriate
variables near the plate, in terms of the Reynolds number ($Re_w=
\overline{V_m}Z_w/\nu$) and Froude number ($Fr_w^2/\beta\Delta T_w$,
where $Fr_w=\overline{V_m}/\sqrt{g Z_w}$) near the plate remains
invariant for a given fluid~(\ref{rew})(\ref{eq:frinv}).

We gratefully acknowledge the partial financial support by DST, Govt.
of India under the grants SR/FST/ETII-017/2003 and SR/S3/MERC/028/2009
for this study. The experiments at $Pr=0.74$ were conducted at DLR,
Gottingen; we acknowledge the financial support of STAR programme of
DAAD and IITM for these experiments. We are grateful to D. Schmeling,
J. Bosbach, O. Shishkina and C. Wagner for their hospitality at
Gottingen. BAP is grateful to J.H. Arakeri for his guidance for the
experiments at $Sc=602$, that were conducted at IISc, Bangalore.
\bibliographystyle{jfm}
\bibliography{guna}
\end{document}